\documentclass[fleqn,usenatbib]{mnras}
\usepackage[T1]{fontenc}
\usepackage{ae,aecompl}
\usepackage{graphicx}
\usepackage{multicol}
\usepackage{float}
\usepackage{longtable}
\usepackage[figuresright]{rotating}

\usepackage{graphicx}	
\usepackage{amsmath}	
\usepackage{amssymb}	
\usepackage{threeparttable}
\usepackage{subfigure}
\usepackage{multirow}
\usepackage{multicol}
\usepackage{booktabs}
\usepackage{makecell}
\newcommand{\orcid}[1]{\href{https://orcid.org/#1}{\includegraphics[width=8pt]{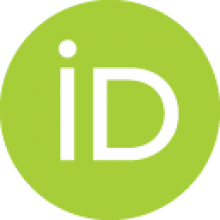}}}

\title[]{Using X-ray continuum-fitting to estimate the spin of MAXI J1305-704}

\author[Ye Feng et al.]{
Ye Feng\orcid{0000-0002-6961-8082}$^{1,2,3}$\thanks{E-mail: yefeng@nao.cas.cn; ye.feng@cfa.harvard.edu},
James F. Steiner\orcid{0000-0002-5872-6061}$^{1}$,
Santiago Ubach Ramirez$^{1}$,
Lijun Gou\orcid{0000-0003-3057-5860}$^{2,3}$
\\
$^{1}$Harvard-Smithsonian Center for Astrophysics, 60 Garden Street, Cambridge, MA 02138, USA\\
$^{2}$Key Laboratory for Computational Astrophysics, National Astronomical Observatories, Chinese Academy of Sciences, \\Datun Road A20, Beijing 100012\\
$^{3}$School of Astronomy and Space Sciences, University of Chinese Academy of Sciences, Datun Road A20, Beijing 100049, China\\
\\
}

\date{Accepted XXX. Received YYY; in original form ZZZ}

\pubyear{2022}

\begin{document}
\pagerange{\pageref{firstpage}--\pageref{lastpage}}
\maketitle

\begin{abstract}
MAXI J1305-704 is a transient X-ray binary with a black hole primary. It was discovered on April 9, 2012, during its only known outburst. MAXI J1305-704 is also a high inclination low-mass X-ray binary with prominent dip features in its light curves, so we check the full catalog of 92 \emph{Swift}/XRT continuous observations of MAXI J1305-704, focusing only on the stable spectra. We select 13 ``gold" spectra for which the root mean square RMS <0.075 and the coronal scattered fraction $f_{\mathrm{sc}} \lesssim 25 \%$. These ``gold" data are optimal thermal-state observations for continuum-fitting modeling, in which the disk extends to the innermost-stable circular orbit and is geometrically thin. The black hole spin was unknown for this object before. By utilizing the X-ray continuum fitting method with the relativistic thin disk model \texttt{kerrbb2} and supplying the known dynamical binary system parameters, we find MAXI J1305-704 has a moderate spin ($a_{*}=0.87_{-0.13}^{+0.07}$) at a 68.3\% confidence level. This is the first determination of MAXI J1305-704's spin.
\end{abstract}


\section{Introduction}
\label{section:1}
The incredible `no-hair theorem' reveals that an astrophysical BH is completely describable by just two quantities: mass and spin. Spin, which can be described by a dimensionless parameter $a_{*} = cJ/GM^2$. Generally, there are two mainstream spectroscopic approaches to constrain the spin of the black hole: the X-ray reflection fitting method (\citealt{Fabian1989}) and the X-ray continuum-fitting (CF) method (\citealt{Zhang1997}). The reflection method has been applied to numerous black holes residing in X-ray binary (BHXRB) systems, e.g., XTE J1550-564 (\citealt{Steiner2011}), GRS 1915+105 (\citealt{Miller2013}), GS 1354-645 (\citealt{El-Batal2016}), MAXI J1535-571 (\citealt{Xu2018}), XTE J1752-223 (\citealt{Javier2018}), MAXI J1836-194 (\citealt{Morningstar2014}; \citealt{Dong2020}), EXO 1846-031(\citealt{Draghis2020}), and CF has been widely applied, e.g., GRS 1915+105 (\citealt{McClintock2006}; \citealt{Middleton2006}), M33 X-7 (\citealt{Liu2008}), A 0620-00 (\citealt{Gou2010}), XTE J1550-564 (\citealt{Steiner2011}), Cyg X-1 (\citealt{Gou2014}), GS 1124-683 (\citealt{Morningstar2014b}; \citealt{Chen2016}).

In the CF method, the inner disk radius ($R_{\rm in}$) is measured by fitting the thermal continuum emission of the accretion disk, which is based on the assumptions of the classical relativistic thin disk model (\citealt{Novikov1973}). By assuming that the inner radius of the accretion disk extends to the innermost stable circular orbit (ISCO), we can estimate the spin using the unique mapping between ISCO and spin discovered in \cite{Bardeen1972}. The CF approach generally requires independent knowledge of system properties, black hole mass ($M$), the source distance ($D$), and accretion disk inclination ($i$, often requiring an assumption that the spin of a black hole is aligned with the orbital angular momentum). The X-ray reflection fitting method also measures the inner disk radius by reprocessing coronal emission in the surface of the disk, especially the Fe K$\alpha$ emission complex. Notably, CF is most reliable in soft/thermal states in which coronal emission is weakest whereas reflection methods are most useful in hard or intermediate states where the coronal emission is strong or dominant. In reflection modeling, the inclination may be fitted in addition to the spin. The CF method is typically used for stellar-mass black holes, whereas the reflection fitting method is widely used for both stellar-mass and supermassive black holes. Apart from the mutual reliance on the relationship between the inner disk radius and the ISCO, both methods are independent and can be used to cross-check each other. We use the CF method to estimate the spin of MAXI J1305-704 in this paper.

Of the approximately 70 black holes and black hole candidate systems detected so far, only five systems are wind-fed and persistent (Cygnus X-1, LMC X-1, M33 X-7, NGC 300 X-1, and IC 10 X-1), while the rest are transient. MAXI J1305-704 is a member of the latter group. It is a Galactic (RA=196.7$^{\circ}$, Dec=-70.5$^{\circ}$) black hole X-ray binary with a peak flux of 30 mCrab (1 Crab\footnote{\url{https://heasarc.gsfc.nasa.gov/docs/heasarc/headates/brightest.html\#compare}} = $2.4 \times 10^{-8}$ $\rm erg s^{-1} cm^{-2}$ for 2.0-10.0 keV) in the 2.0-4.0 keV band (\citealt{Sato2012}). MAXI J1305-704 was first noticed by the Monitor of an All-sky X-ray Image/Gas Slit Camera (\emph{MAXI}/GSC) (\citealt{Sato2012}) on 2012 April 9 at 11:24:23 UT. Multiple wavelength analyses indicated that the source is most likely a BHXRB (\citealt{Greiner2012}; \citealt{Kennea2012a}; \citealt{Suwa2012}; \citealt{Charles2012}; \citealt{Morihana2013}). It exhibited canonical hard, intermediate, and soft states during the outburst (\citealt{Suwa2012}; \citealt{Morihana2013}). A ground-based dynamical study performed in quiescence later confirmed that MAXI J1305-704 is a black hole with the mass of $M_{\rm BH}=8.9_{-1.0}^{+1.6} M_{\odot}$, a companion mass of $M_{\mathrm{opt}}=0.43 \pm 0.16 M_{\odot}$, at a distance of $D=7.5_{-1.4}^{+1.8} \mathrm{kpc}$, and a high inclination of $i={72_{-8}^{+5}}^{\circ}$ (uncertainties are 1 $\sigma$; \citealt{Mata2021}). From these data, we are able to apply CF and measure its spin. Previous studies have shown that MAXI J1305-704 is an anomalous source, showing a very high disk temperature ($\sim$1.0 keV) in the soft state, even while at extremely low luminosity ($\sim$0.02 $L_{\mathrm{Edd}}$) (\citealt{Shidatsu2013}; \citealt{Morihana2013}). In general, black holes with such high disk temperatures usually have relatively large luminosity.

Using the \cite{Mata2021} dynamical results, we apply the relativistic thin-disk model \texttt{kerrbb2} (\citealt{McClintock2006}; which merges two different disk models, \texttt{kerrbb} (\citealt{Li2005}) and \texttt{bhspec} (\citealt{Davis2005})) with CF approach to fit an optimal selection of 13 ``gold" spectra (13 selected spectra of \emph{Swift}/XRT in the soft state during its 2012 outburst) so as to further constrain its spin. This is because, in the soft state, the inner radius of the accretion disk has extended to ISCO, and at sufficiently low luminosities the accretion disk conforms to a standard thin disk structure.

The structure of the paper is as follows. We present data selection and reduction in Section \ref{section:2}. Section \ref{section:3} delves deeply into the spectral analysis and results. Sections \ref{section:4} and \ref{section:5} present the discussions and conclusion respectively.

\section{DATA SELECTION AND REDUCTION} 
\label{section:2}
From \emph{MAXI}/GSC's thirteen years of coverage, MAXI J1305-704 exhibited only one significant outburst in 2012 as shown in Figure \ref{fig:1_1}, nor was it detected by predecessor X-ray sky monitors (e.g., \emph{Swift}/BAT or \emph{RXTE}/ASM). The corresponding hardness-intensity diagram (HID) in Figure \ref{fig:1_2} exhibits a non-standard shape, distinct from the usual ``q"-like profile typical of major outbursts by BHXRBs. During this outburst in 2012, \emph{Swift}/XRT (\citealt{Burrows2005}) had two observing programs (00032339XXX and 00032461XXX), totaling 65 observations containing 92 continuous data segments. Consistent with its high orbital inclination (and similar to other high-inclination BHXRBs), MAXI J1305-704 exhibited obvious dip characteristics in its light-curves (e.g., \citealt{Kennea2012c}; \citealt{Kennea2012b}; \citealt{Shidatsu2013}; \citealt{Miller2014}). Regarding the 65 observations by \emph{Swift}/XRT, we used the FTOOLS\footnote{\url{http://heasarc.gsfc.nasa.gov/docs/software/}} command \texttt{ftselect}\footnote{\url{https://heasarc.gsfc.nasa.gov/lheasoft/ftools/headas/ftselect.html}} to divide them by Good Time Interval (GTI), into 92 spectra and light curves, which are summarized in Table \ref{table:1}. Figure \ref{fig:2} shows the hardness evolution over time. Figure \ref{fig:3} depicts three types of features of the light curve for illustration. In general, the spectra used to measure the spin should have no dips, flares, or other sizeable variations. Each light curve is screened by eye to exclude data sets with dips or presenting high variability. ``Dip" light curves (Figure \ref{fig:3_1}) display one or more sharply-shaped deficits in count rate, whereas ``variable" light curves (Figure \ref{fig:3_2}) show significant variations that are inconsistent with a stable and fixed count rate. To ensure robustness, spectra are required to have at least 8000 counts. We also screen for spectra whose emission is dominated by the accretion disk (\citealt{McClintock2006}) since the CF technique relies upon the signal of the thermal disk component, and also presumes that the inner radius of the disk reaches the ISCO, which is most firmly grounded for soft disk-dominated states (e.g., \citealt{Tanaka1995}; \citealt{Kubota2003}; \citealt{Steiner2010}). This also eliminates the possibility of contamination by a strong Compton component. We adopt an upper limit on the root mean square (RMS) variability (RMS < 0.075, \citealt{Remillard2006}) and the relative strength of the non-thermal emission ($f_{\mathrm{sc}} \lesssim 25 \%$, where $f_{\mathrm{sc}}$ is the relative strength of the Comptonization model \texttt{simpl} in \texttt{XSPEC}. See \citealt{Steiner2009}; \citealt{Steiner2009b}). More details are given in Section \ref{section:3.1}.

\setcounter{figure}{0}
\begin{figure}
\subfigure{
    \label{fig:1_1}
    \includegraphics[angle=0, width=\columnwidth]{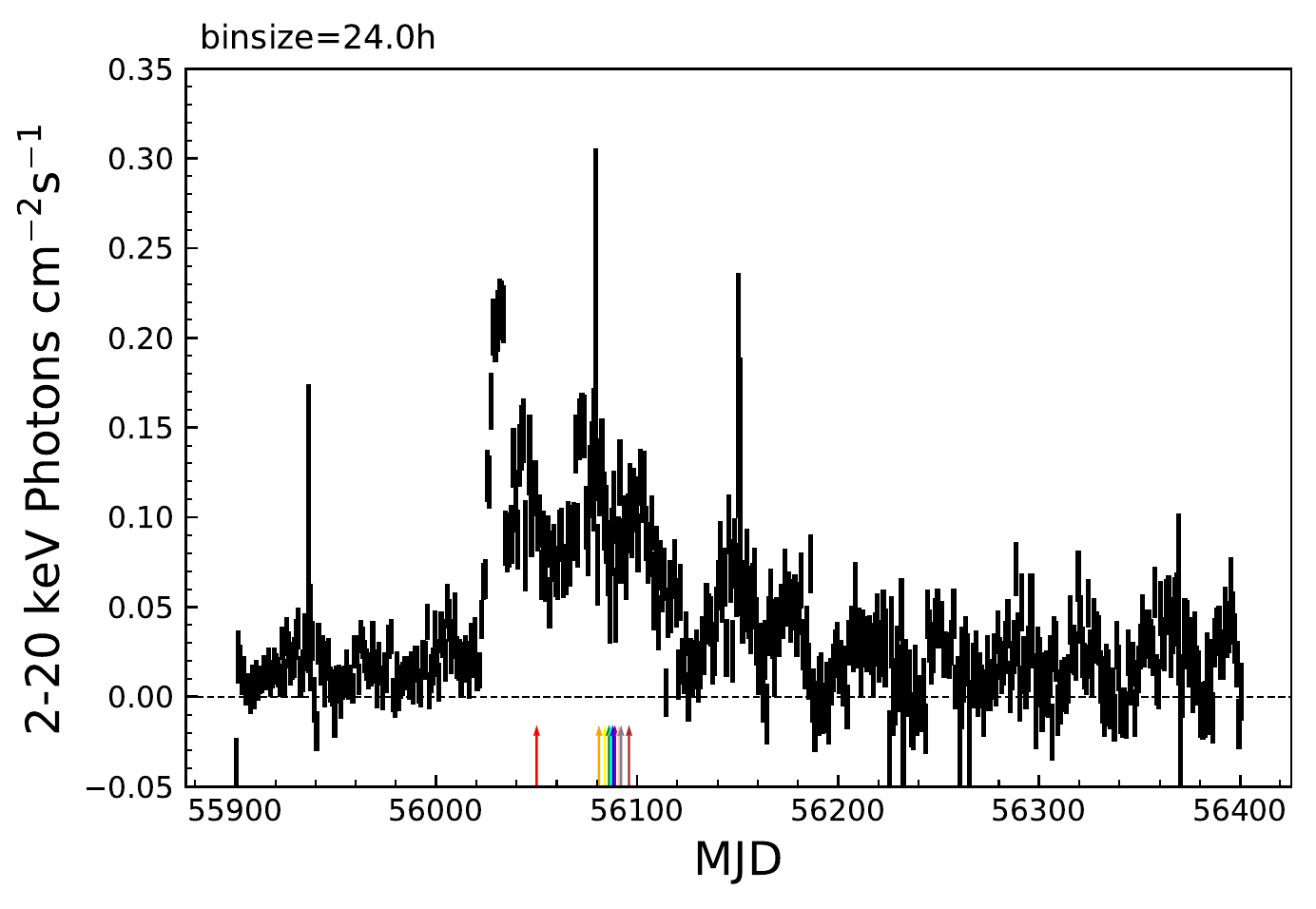}
    }
    \subfigure{
    \includegraphics[angle=0, width=\columnwidth]{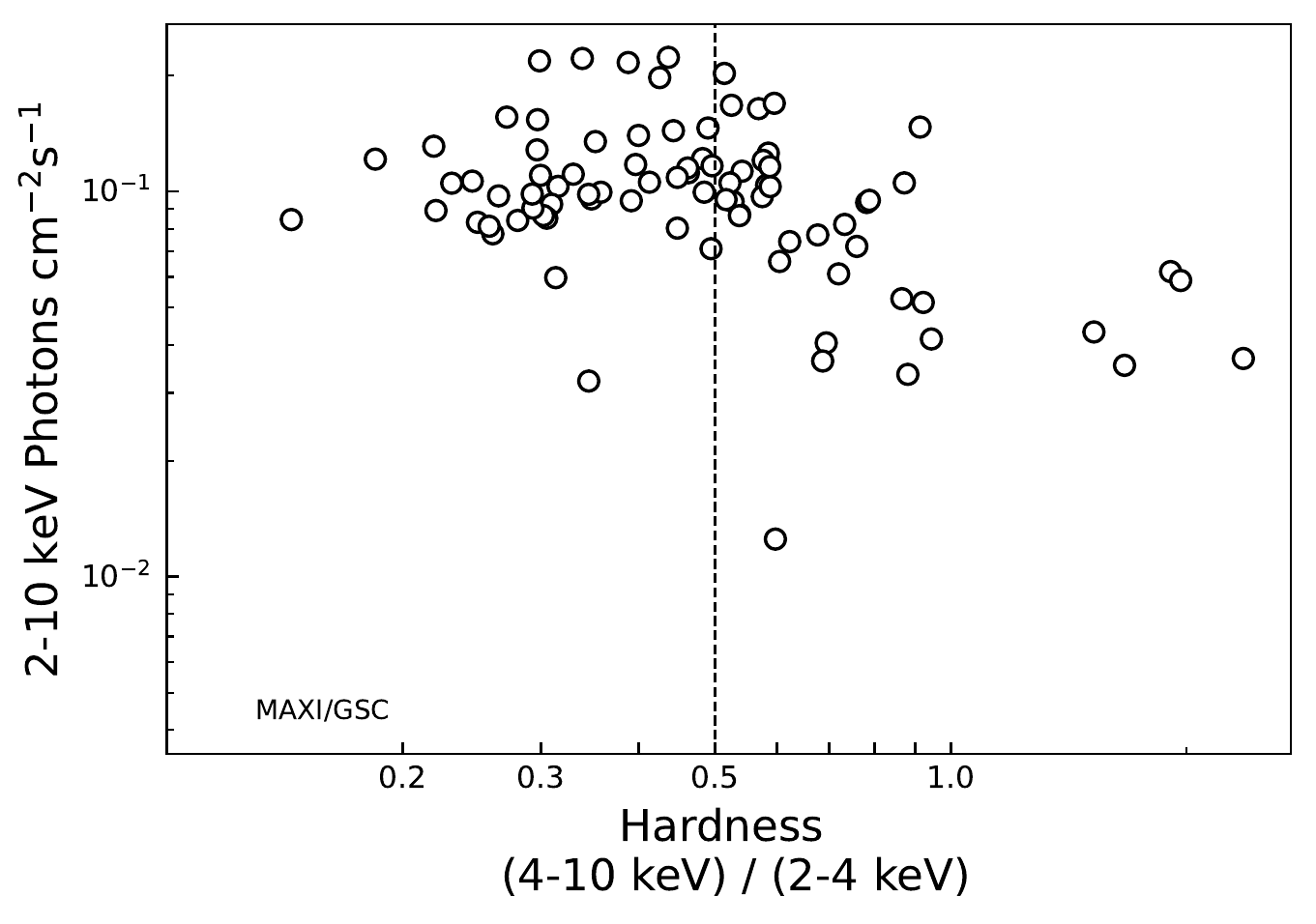}
      \label{fig:1_2}
    }
    \caption{a) The light curve of the MAXI J1305-704 outburst was observed by \emph{MAXI}/GSC in 2012. The colored arrows mark the dates of SP1 through SP13. b) The HID of MAXI J1305-704 is based on \emph{MAXI}/GSC measurements. It corresponds to the outburst from MJD=56022 to MJD=56120 in Figure \ref{fig:1_1}. The \emph{MAXI}/GSC hardness is expressed as the ratio of counts measured at 4.0-10.0 keV to counts detected at 2.0-4.0 keV. A dashed line at a hardness value of 0.5 is included as a visual reference.}
\end{figure}

\setcounter{figure}{1}
\begin{figure}
    \includegraphics[angle=0, width=\columnwidth]{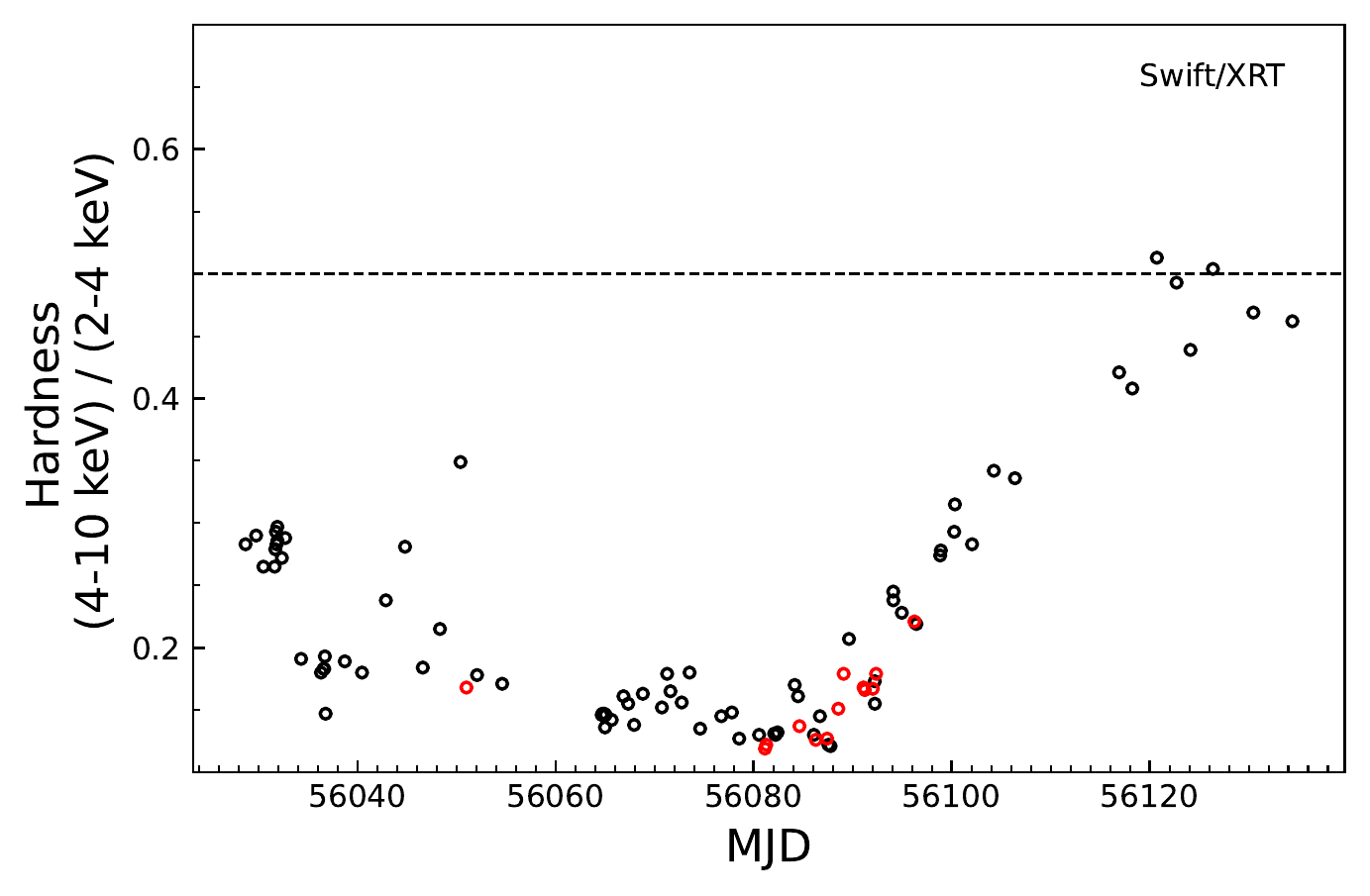}
    \caption{The hardness over time for 92 spectra of \emph{Swift}/XRT, using 4.0-10.0 keV/2.0-4.0 keV rates, but with different amplitude than the \emph{MAXI}/GSC values in Figure \ref{fig:1_2} owing to differences in the instruments. SP1-SP13 are marked in red. A dashed line at a hardness value of 0.5 is included as a visual reference.}\label{fig:2}
\end{figure}

\setcounter{figure}{2}
\begin{figure}
\subfigure{
    \label{fig:3_1}
    \includegraphics[angle=0, width=\columnwidth]{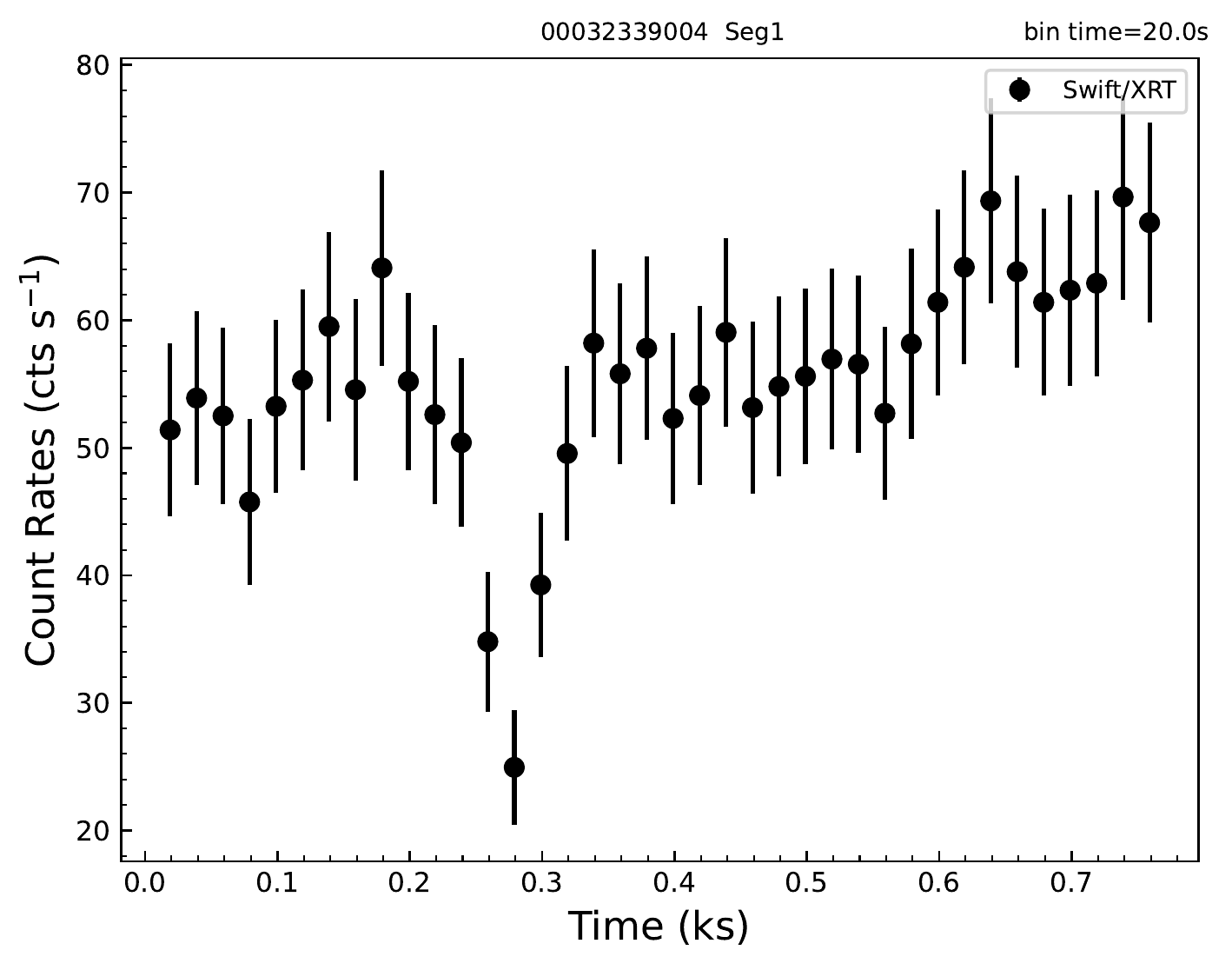}
    }
\subfigure{
    \includegraphics[angle=0, width=\columnwidth]{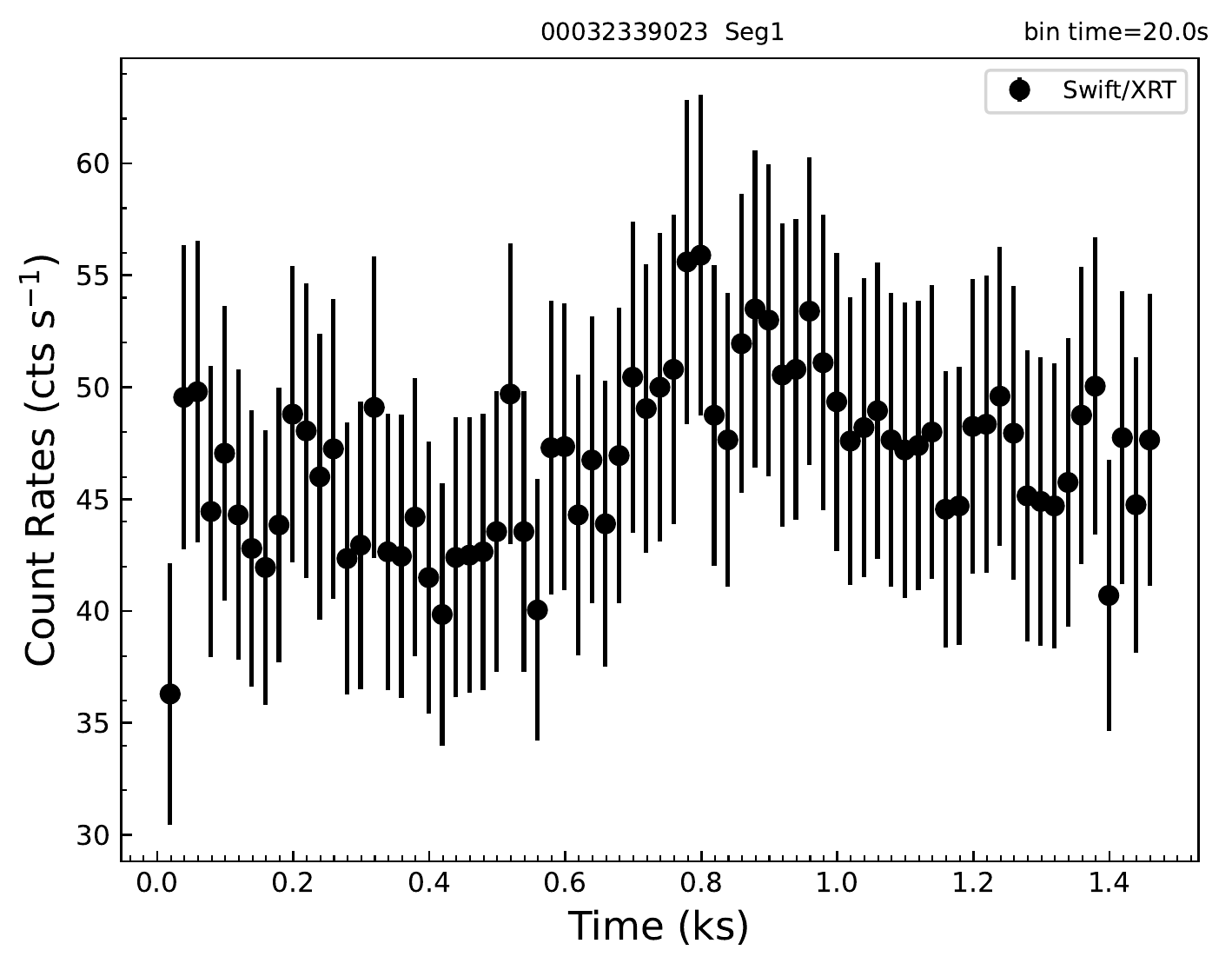}
      \label{fig:3_2}
    }
\subfigure{
    \label{fig:3_3}
    \includegraphics[angle=0, width=\columnwidth]{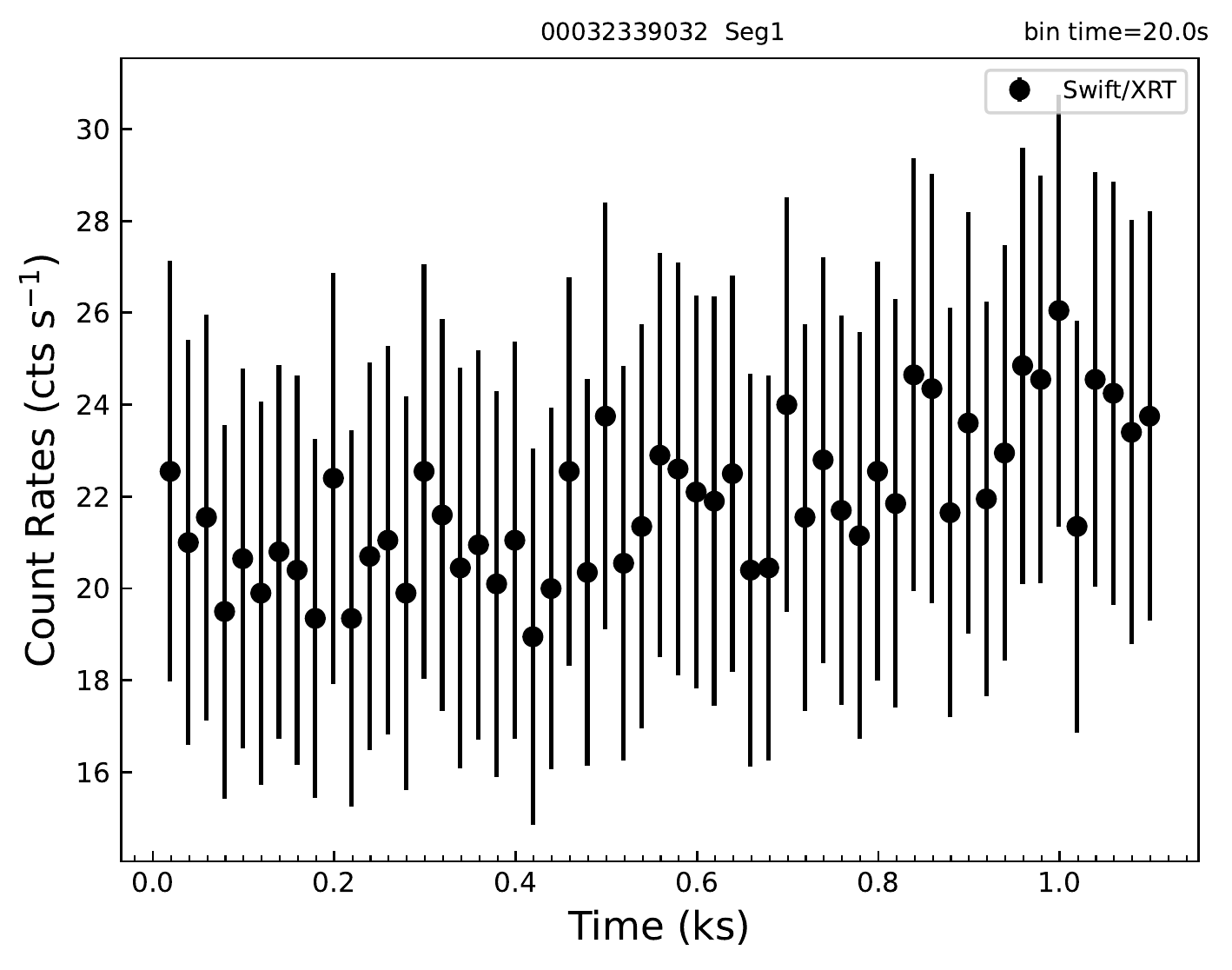}
    }
    \caption{a) N9 (ObsID: 00032339004 Seg1), an observation of \emph{Swift}/XRT, where the net light curve generated at 0.3-10.0 keV displays a noticeable dip. b) N43 (ObsID: 00032339023 Seg1), an observation of \emph{Swift}/XRT, where the net light curve generated at 0.3-10.0 keV shows a significant variation. c) N69 (ObsID: 00032339032 Seg1), an observation of \emph{Swift}/XRT, where the stable net light curve generated at 0.3-10.0 keV.}\label{fig:3}
\end{figure}

\setcounter{figure}{3}
\begin{figure} 
    \includegraphics[angle=0, width=\columnwidth]{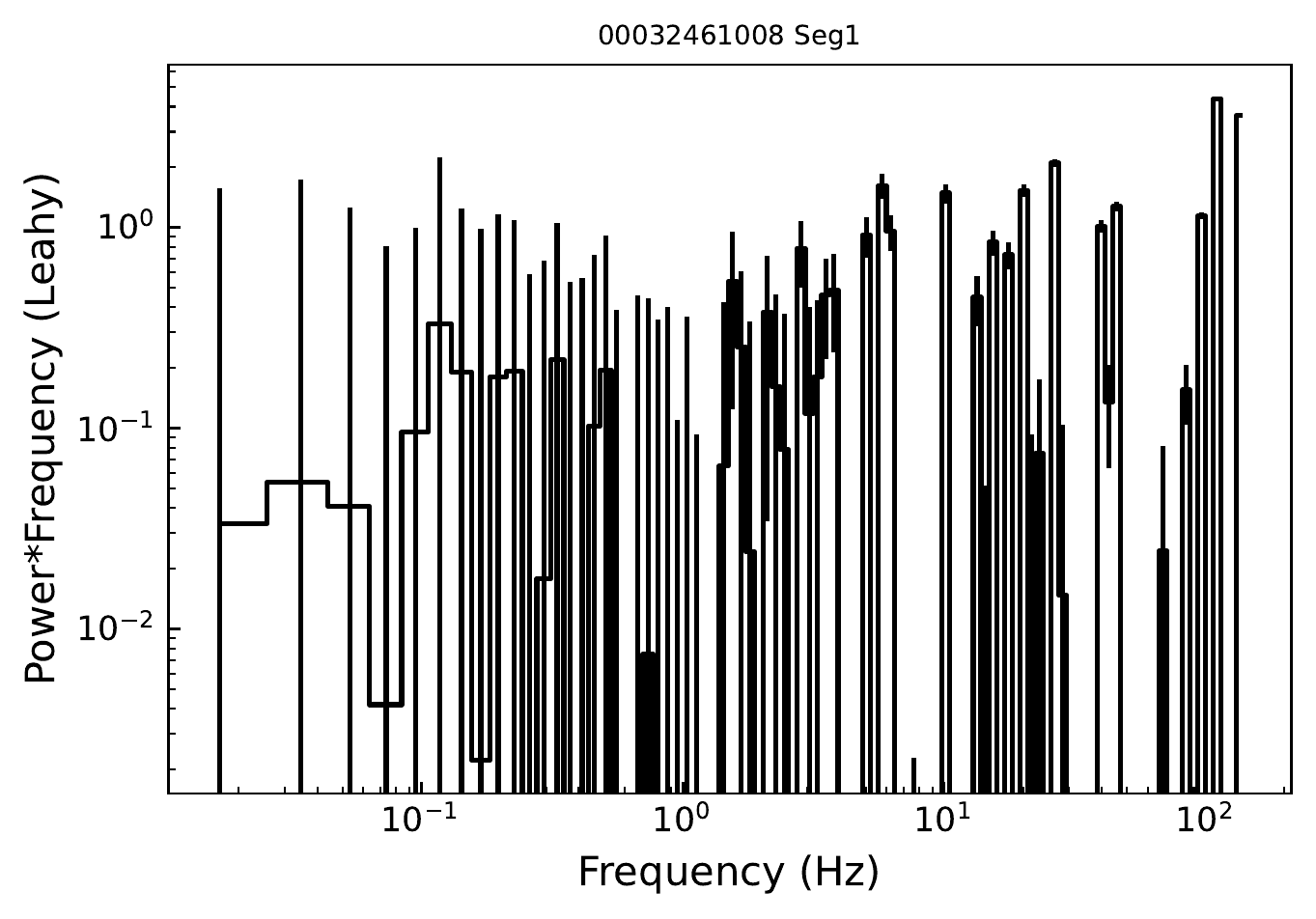}
    \caption{The PSD plot (with Leahy normalization while having subtracted off the Poisson noise) for N48 (SP2) (ObsId:00032461008 Seg1), which is representative of the other thermal data. The RMS value is $0.011 \pm 0.007$, and the frequency range is from 0.017 to 139 Hz.}\label{fig:4}
\end{figure}

\begin{table*}
   \renewcommand\arraystretch{0.8} 
    \centering
	 \caption{MAXI J1305-704 Swift/XRT observation overview}
    \begin{threeparttable}
    \resizebox{\textwidth}{!}{
    \begin{tabular}{ccccccccccc}
        \hline
        \hline
Number & ObsID & Seg. & MJD  &  Start time  & Exposure & Count Rates & Phase & RMS & LC type & $f_{\rm sc}$ \\
 &  & &   &    & (s) & (cts $\rm s^{-1}$) &  &  &  &  \\
\hline         
N1  &  00032339001 &  Seg1    &  56028  &  2012-04-11 16:08:46.801  &  997        &  44.27     &  0.000-0.029 &  0.137  $\pm$ 0.005       &variable     &-\\ 
N2  &  00032339002 &  Seg1    &  56029  &  2012-04-12 17:54:32.401  &  1282       &  46.67     &  0.718-0.755 &  0.084  $\pm$ 0.004       & dips              &-\\   
N3  &  00032339003 &  Seg1    &  56030  &  2012-04-13 11:30:40.601  &  858        &  55.14     &  0.574-0.600 &  0.094  $\pm$ 0.005       &variable     &-\\  
N4  &  00032339006 &  Seg1    &  56031  &  2012-04-14 14:42:48.001  &  2418       &  51.83     &  0.444-0.515 &  0.110  $\pm$ 0.003       &dips               &-\\  
N5  &  00032339006 &  Seg2    &  56031  &  2012-04-14 16:18:03.601  &  2474       &  59.60     &  0.611-0.684 &  0.086  $\pm$ 0.003       &dips               &-\\   
N6  &  00032339006 &  Seg3    &  56031  &  2012-04-14 17:54:44.801  &  2443       &  29.64     &  0.781-0.853 &  0.146  $\pm$ 0.004       &dips               &-\\ 
N7  &  00032339006 &  Seg4    &  56031  &  2012-04-14 19:29:49.201  &  2513       &  37.40     &  0.948-0.023 &  0.071  $\pm$ 0.003       &dips               &-\\
N8  &  00032339006 &  Seg5    &  56031  &  2012-04-14 21:23:23.201  &  129        &  56.40     &  0.148-0.152 &  0.035  $\pm$ 0.012       &stable             &-\\
N9  &  00032339004 &  Seg1    &  56031  &  2012-04-14 21:26:11.801  &  763        &  55.51     &  0.153-0.175 &  0.128  $\pm$ 0.005       &dips                 &-\\ 
N10 &  00032339005 &  Seg1    &  56032  &  2012-04-15 08:23:37.001  &  683        &  53.12     &  0.309-0.330 &  0.088  $\pm$ 0.005       &stable             &-\\ 
N11 &  00032339005 &  Seg2    &  56032  &  2012-04-15 16:23:24.801  &  693        &  61.06     &  0.152-0.173 &  0.066  $\pm$ 0.005       &variable     &-\\ 
N12 &  00032339007 &  Seg1    &  56034  &  2012-04-17 06:58:08.401  &  1068       &  36.31     &  0.222-0.253 &  0.169  $\pm$ 0.005       &dips               &-\\ 
N13 &  00032339008 &  Seg1    &  56036  &  2012-04-19 07:09:15.201  &  943        &  36.17     &  0.305-0.332 &  0.088  $\pm$ 0.005      &dips                 &-\\      
N14 &  00032339009 &  Seg1    &  56036  &  2012-04-19 13:19:54.201  &  2459       &  20.50     &  0.956-0.028 &  0.166  $\pm$ 0.004       &dips               &-\\
N15 &  00032339009 &  Seg2    &  56036  &  2012-04-19 14:56:45.401  &  2419       &  30.42     &  0.127-0.198 &  0.125  $\pm$ 0.004       &dips               &-\\
N16 &  00032339009 &  Seg3    &  56036  &  2012-04-19 16:30:53.601  &  2543       &  33.27     &  0.292-0.367 &  0.165  $\pm$ 0.003       &dips               &-\\
N17 &  00032339009 &  Seg4    &  56036  &  2012-04-19 18:18:42.201  &  87         &  41.50     &  0.482-0.485 &                 < 0.020            &stable             &-\\ 
N18 &  00032339009 &  Seg5    &  56038  &  2012-04-21 16:46:57.201  &  958        &  45.73     &  0.384-0.412 &  0.076  $\pm$ 0.005       &variable     &-\\
N19 &  00032339010 &  Seg1    &  56040  &  2012-04-23 10:34:56.201  &  470        &  43.38     &  0.793-0.807 &  0.155  $\pm$ 0.007       & dips               &-\\ 
N20 &  00032339011 &  Seg1    &  56042  &  2012-04-25 20:24:11.601  &  943        &  24.74     &  0.892-0.920 &  0.115  $\pm$ 0.007       & dips                &-\\
N21 &  00032339012 &  Seg1    &  56044  &  2012-04-27 18:53:20.201  &  818        &  25.03     &  0.796-0.820 &  0.177  $\pm$ 0.007      & dips                &-\\  
N22 &  00032339013 &  Seg1    &  56046  &  2012-04-29 13:51:04.601  &  1013       &  50.78     &  0.328-0.357 &  0.070  $\pm$ 0.004       &variable    &-\\ 
N23 &  00032339014 &  Seg1    &  56048  &  2012-05-01 07:31:04.801  &  893        &  29.63     &  0.723-0.749 &                 < 0.006       &variable         &-\\
N24 &  00032339015 &  Seg1    &  56050  &  2012-05-03 09:17:51.601  &  667        &  5.31      &  0.974-0.994 &  0.378  $\pm$ 0.017       &variable      &-\\
N25 &  00032339015 &  Seg2    &  56050  &  2012-05-03 23:39:55.001  &  723        &  43.76     &  0.489-0.511 &  0.062  $\pm$ 0.006       &stable              &0.098 $\pm$ 0.012\\ 
N26 &  00032339016 &  Seg1    &  56052  &  2012-05-05 01:26:51.201  &  967        &  17.08     &  0.209-0.237 &  0.163  $\pm$ 0.008       &dips                &-\\ 
N27 &  00032339017 &  Seg1    &  56054  &  2012-05-07 14:21:16.601  &  997        &  35.85     &  0.634-0.663 &  0.057  $\pm$ 0.005       &variable     &-\\ 
N28 &  00032461001 &  Seg1    &  56064  &  2012-05-17 15:41:23.001  &  696        &  32.83     &  0.091-0.112 &  0.088  $\pm$ 0.007       &variable    &-\\ 
N29 &  00032461001 &  Seg2    &  56064  &  2012-05-17 17:17:23.001  &  753        &  35.77     &  0.260-0.282 &  0.123  $\pm$ 0.006       &variable     &-\\ 
N30 &  00032339018 &  Seg1    &  56064  &  2012-05-17 22:02:22.801  &  457        &  30.29     &  0.761-0.775 &  0.086  $\pm$ 0.009       &variable     &-\\  
N31 &  00032339018 &  Seg2    &  56064  &  2012-05-17 23:36:47.801  &  488        &  32.77     &  0.927-0.941 &  0.083  $\pm$ 0.008       &variable     &-\\ 
N32 &  00032461002 &  Seg1    &  56065  &  2012-05-18 01:12:16.601  &  463        &  31.85     &  0.095-0.109 &  0.113  $\pm$ 0.009       &variable     &-\\ 
N33 &  00032461002 &  Seg2    &  56065  &  2012-05-18 15:56:06.201  &  174        &  24.80     &  0.649-0.654 &  0.098  $\pm$ 0.018       &variable    &-\\
N34 &  00032339019 &  Seg1    &  56066  &  2012-05-19 20:24:52.801  &  1023       &  34.79     &  0.653-0.683 &  0.132  $\pm$ 0.005       &variable    &-\\
N35 &  00032461003 &  Seg1    &  56067  &  2012-05-20 07:45:17.001  &  343        &  35.13     &  0.849-0.860 &                 < 0.010            &variable    &-\\
N36 &  00032461003 &  Seg2    &  56067  &  2012-05-20 22:11:19.001  &  518        &  38.51     &  0.372-0.387 &  0.100  $\pm$ 0.007       &variable    &-\\
N37 &  00032461004 &  Seg1    &  56068  &  2012-05-21 19:20:13.001  &  164        &  43.75     &  0.602-0.607 &  0.066  $\pm$ 0.014       &variable     &-\\ 
N38 &  00032339021 &  Seg1    &  56070  &  2012-05-23 17:32:14.601  &  984        &  19.07     &  0.476-0.505 &  0.151  $\pm$ 0.007       &stable             &-\\ 
N39 &  00032461005 &  Seg1    &  56071  &  2012-05-24 06:12:02.601  &  717        &  47.56     &  0.812-0.833 &  0.084  $\pm$ 0.005       &variable     &-\\ 
N40 &  00032461005 &  Seg2    &  56071  &  2012-05-24 14:10:32.001  &  443        &  44.02     &  0.653-0.666 &                 <0.007            &variable     &-\\ 
N41 &  00032339022 &  Seg1    &  56072  &  2012-05-25 17:33:18.601  &  1417       &  47.67     &  0.541-0.583 &  0.073  $\pm$ 0.004       &variable    &-\\ 
N42 &  00032461006 &  Seg1    &  56073  &  2012-05-26 12:38:41.801  &  1452       &  50.33     &  0.555-0.597 &                 <0.004            &variable    &-\\ 
N43 &  00032339023 &  Seg1    &  56074  &  2012-05-27 14:15:20.601  &  1477       &  46.89     &  0.256-0.300 &                 < 0.004           &variable    &-\\  
N44 &  00032339024 &  Seg1    &  56076  &  2012-05-29 17:35:42.401  &  1452       &  48.56     &  0.672-0.714 &  0.080  $\pm$ 0.004       &variable    &-\\ 
N45 &  00032461007 &  Seg1    &  56077  &  2012-05-30 19:15:30.801  &  1407       &  27.53     &  0.379-0.420 &  0.077  $\pm$ 0.005       &variable    &-\\ 
N46 &  00032339025 &  Seg1    &  56078  &  2012-05-31 12:55:51.401  &  966        &  20.10     &  0.243-0.272 &  0.136  $\pm$ 0.007       &stable             &-\\ 
N47 &  00032339026 &  Seg1    &  56080  &  2012-06-02 13:06:25.401  &  933        &  29.45     &  0.325-0.352 &  0.098  $\pm$ 0.006       &variable    &-\\  
N48 &  00032461008 &  Seg1    &  56081  &  2012-06-03 03:20:06.801  &  532        &  42.53     &  0.826-0.842 &  0.011  $\pm$ 0.007       &stable             &0.051 $\pm$ 0.010 \\ 
N49 &  00032461008 &  Seg2    &  56081  &  2012-06-03 06:32:06.201  &  473        &  44.48     &  0.163-0.177 &                   < 0.007         & stable            &0.072 $\pm$ 0.011\\ 
N50 &  00032461009 &  Seg1    &  56082  &  2012-06-04 01:47:07.201  &  712        &  30.20     &  0.194-0.216 &                    < 0.007       &  dips               &-\\
N51 &  00032339027 &  Seg1    &  56082  &  2012-06-04 05:06:45.401  &  1085       &  30.65     &  0.545-0.577 &  0.098  $\pm$ 0.006       &variable    &-\\
N52 &  00032461009 &  Seg2    &  56082  &  2012-06-04 09:57:14.401  &  643        &  34.63     &  0.056-0.075 &                    < 0.007       &variable      &-\\
N53 &  00032461010 &  Seg1    &  56084  &  2012-06-06 03:27:44.601  &  1334       &  20.09     &  0.434-0.474 &  0.091  $\pm$ 0.006       &variable  &-\\
N54 &  00032461010 &  Seg2    &  56084  &  2012-06-06 11:28:44.401  &  1153       &  49.86     &  0.280-0.314 &                     < 0.004       &  dips              &-\\ 
N55 &  00032339028 &  Seg1    &  56084  &  2012-06-06 15:01:15.201  &  1003       &  46.42     &  0.653-0.683 &  0.023  $\pm$ 0.005       &stable            &0.076 $\pm$ 0.007 \\ 
N56 &  00032461011 &  Seg1    &  56086  &  2012-06-08 01:58:19.201  &  1001       &  36.23     &  0.340-0.370 &  0.062  $\pm$ 0.005       &variable   &-\\ 
N57 &  00032461011 &  Seg2    &  56086  &  2012-06-08 06:45:19.201  &  938        &  41.04     &  0.845-0.872 &                 < 0.005            &stable            &0.064 $\pm$ 0.007\\
N58 &  00032339029 &  Seg1    &  56086  &  2012-06-08 16:35:17.801  &  997        &  42.98     &  0.882-0.911 &  0.036  $\pm$ 0.005       &variable     &-\\ 
N59 &  00032461012 &  Seg1    &  56087  &  2012-06-09 10:00:19.201  &  820        &  44.80     &  0.719-0.744 &                < 0.005            &stable             &0.059 $\pm$ 0.007\\ 
N60 &  00032461012 &  Seg2    &  56087  &  2012-06-09 13:15:19.201  &  640        &  46.96     &  0.062-0.082 &                < 0.006             &variable   &-\\ 
N61 &  00032461012 &  Seg3    &  56087  &  2012-06-09 18:22:07.401  &  768        &  45.68     &  0.602-0.624 &                < 0.005            &dips               &-\\
N62 &  00032339030 &  Seg1    &  56088  &  2012-06-10 13:15:19.401  &  937        &  18.05     &  0.594-0.621 &                < 0.008           &stable              &0.134 $\pm$ 0.014  \\
N63 &  00032461013 &  Seg1    &  56089  &  2012-06-11 02:05:19.401  &  1239       &  45.47     &  0.948-0.985 &                < 0.004           &stable              &0.181 $\pm$ 0.009  \\
N64 &  00032461013 &  Seg2    &  56089  &  2012-06-11 15:04:04.601  &  773        &  50.93     &  0.317-0.339 &                 < 0.005           &variable      &-\\
N65 &  00032461014 &  Seg1    &  56091  &  2012-06-13 02:20:06.401  &  892        &  48.86     &  0.037-0.064 &                 <0.005           &stable              &0.135 $\pm$ 0.009 \\
N66 &  00032461014 &  Seg2    &  56091  &  2012-06-13 05:28:54.001  &  1083       &  51.42     &  0.369-0.400 &                <0.004            &stable              &0.147 $\pm$ 0.008 \\
N67 &  00032461015 &  Seg1    &  56092  &  2012-06-14 00:44:28.201  &  631        &  20.09     &  0.400-0.420 &                 < 0.009          &stable               &0.126 $\pm$ 0.015\\ 
N68 &  00032461015 &  Seg2    &  56092  &  2012-06-14 05:32:28.201  &  572        &  48.18     &  0.907-0.924 &                 < 0.006          &variable      &-\\
N69 &  00032339032 &  Seg1    &  56092  &  2012-06-14 05:42:24.601  &  1113       &  21.92     &  0.924-0.957 &  0.082  $\pm$ 0.007        & stable            &-\\ 
N70 &  00032461015 &  Seg3    &  56092  &  2012-06-14 08:45:28.201  &  568        &  48.88     &  0.246-0.263 &                 < 0.006          &stable              &0.171 $\pm$ 0.012\\
N71 &  00032461016 &  Seg1    &  56094  &  2012-06-16 02:18:34.601  &  1044       &  45.79     &  0.629-0.661 &                < 0.005           &variable     &-\\   
N72 &  00032339033 &  Seg1    &  56094  &  2012-06-16 02:36:17.801  &  1118       &  42.97     &  0.660-0.693 &                 < 0.005          &stable               &0.279 $\pm$ 0.010 \\
N73 &  00032461016 &  Seg2    &  56094  &  2012-06-16 23:07:45.401  &  972        &  43.00     &  0.825-0.854 &                 < 0.005          &variable     &-\\ 
N74 &  00032461017 &  Seg1    &  56096  &  2012-06-18 05:43:47.001  &  971        &  43.80     &  0.053-0.082 &                 < 0.005          &stable              &0.239 $\pm$ 0.011\\  
N75 &  00032339034 &  Seg1    &  56096  &  2012-06-18 07:31:10.201  &  986        &  45.16     &  0.242-0.271 &                  < 0.005         &variable     &-\\
N76 &  00032461017 &  Seg2    &  56096  &  2012-06-18 10:24:45.401  &  1452       &  45.13     &  0.547-0.590 &                < 0.004           &variable       &-\\ 
N77 &  00032339035 &  Seg1    &  56098  &  2012-06-20 20:06:12.601  &  467        &  32.80     &  0.633-0.647 &                  < 0.009         &stable               &0.338 $\pm$ 0.018 \\ 
N78 &  00032339035 &  Seg2    &  56098  &  2012-06-20 21:43:12.601  &  823        &  34.76     &  0.803-0.827 &                  < 0.006         &stable               &0.352 $\pm$ 0.012\\ 
N79 &  00032339036 &  Seg1    &  56100  &  2012-06-22 06:04:03.401  &  1015       &  29.33     &  0.215-0.246 &                 < 0.006         &stable               &0.389 $\pm$ 0.012\\
N80 &  00032339036 &  Seg2    &  56100  &  2012-06-22 07:48:09.001  &  528        &  30.87     &  0.398-0.414 &                  < 0.008         &stable               &0.408 $\pm$ 0.018 \\
N81 &  00032339037 &  Seg1    &  56102  &  2012-06-24 01:19:14.801  &  1003       &  32.52     &  0.778-0.807 &  0.111  $\pm$ 0.006        &stable             &-\\
N82 &  00032339038 &  Seg1    &  56104  &  2012-06-26 06:10:49.801  &  1028       &  22.12     &  0.354-0.384 &                 < 0.007         &stable               &0.437 $\pm$ 0.025\\
N83 &  00032339039 &  Seg1    &  56106  &  2012-06-28 09:20:03.801  &  1013       &  22.30     &  0.750-0.780 &  0.141  $\pm$ 0.007       &stable             &-\\ 
N84 &  00032339040 &  Seg1    &  56116  &  2012-07-08 21:52:16.601  &  818        &  12.17     &  0.389-0.413 &  0.055  $\pm$ 0.010      &stable               &0.589 $\pm$ 0.145\\ 
N85 &  00032339041 &  Seg1    &  56118  &  2012-07-10 05:48:21.201  &  997        &  10.76     &  0.757-0.787 &  0.267  $\pm$ 0.010       &stable             &-\\ 
N86 &  00032339042 &  Seg1    &  56120  &  2012-07-12 17:19:28.601  &  149        &  6.89      &  0.036-0.040 &  0.425  $\pm$ 0.035       &stable               &-\\ 
N87 &  00032339043 &  Seg1    &  56122  &  2012-07-14 17:06:23.201  &  1112       &  4.56      &  0.076-0.109 &  0.420  $\pm$ 0.014       & dips                 &-\\ 
N88 &  00032339044 &  Seg1    &  56124  &  2012-07-16 02:46:06.001  &  1012       &  4.64      &  0.627-0.657 &  0.390  $\pm$ 0.015       &stable              &-\\
N89 &  00032339045 &  Seg1    &  56126  &  2012-07-18 09:18:11.001  &  1007       &  3.03      &  0.379-0.409 &  0.277  $\pm$ 0.018       &dips                  &-\\ 
N90 &  00032339046 &  Seg1    &  56128  &  2012-07-20 09:36:32.001  &  1042       &  1.08      &  0.475-0.506 &  0.687  $\pm$ 0.030       &variable       
    &-\\ 
N91 &  00032339047 &  Seg1    &  56130  &  2012-07-22 11:19:10.801  &  1006       &  3.85      &  0.719-0.748 &  0.469  $\pm$ 0.016       &stable                &-\\ 
N92 &  00032339049 &  Seg1    &  56134  &  2012-07-26 09:53:27.401  &  987        &  3.49      &  0.695-0.724 &  0.472  $\pm$ 0.017       &dips                   &-\\
        \hline
       
    \end{tabular}}
    \begin{tablenotes}
        \footnotesize
    \item Notes. In columns 1-12, we show, in turn: the number of spectra that are split according to GTI; the observation ID (ObsID);\\ the segment number of the current ObsID; Modified Julian Date (MJD); the start time; the exposure time in units of s; the net count rates\\ (0.3-10.0 keV) in units of cts $\rm s^{-1}$; the relative phase for which we arbitrarily choose to reference phase=0 at the start of N1 \\(there is insufficient precision in the period to obtain a robust absolute phasing based on the optical studies), with the orbital \\period $P_{\text {orb }}=0.394 \pm 0.004$ days (\citealt{Mata2021}); the total root mean square (RMS) power integrated between 0.1 and 10.0 Hz\\ in the power spectral density over 0.3-10.0 keV; the features of light-curves, the character of the light curve (stable, variable, or \\exhibiting dips); the scattered fraction ($f_{\rm sc}$), only for a subset who meets the RMS <0.075 and light-curve is stable.
    \end{tablenotes}
    \end{threeparttable}
    \label{table:1}
\end{table*}

\subsection{Spectra}
\label{section:2.1}
All \emph{Swift}/XRT observations are performed in the Windowed Timing (WT) mode (1-dimensional imaging) which can avoid pile-up effects. We utilize grades 0-2 for WT mode. A dead-time correction is taken into account. We reduce the archival data using the command \texttt{xrtpipeline} v0.13.5 in HEAsoft v6.28 with \emph{Swift}/XRT Calibration Database (CALDB) v20210915. The source spectra are extracted following the official recommendations\footnote{\url{https://www.swift.ac.uk/analysis/xrt/xselect.php}} with the command \texttt{xselect} v2.4k\footnote{\url{https://www.swift.ac.uk/analysis/xrt/spectra.php}} from a circular region of 20 pixels (1 pixel =2.36 arcsec), whose center is located at the source position. Background spectra are extracted from an annulus with the inner and outer radii of $80^{\prime \prime}$ and $120^{\prime \prime}$ respectively, centered on the source. We also modify the header keyword \texttt{backscal} of the source and background spectra respectively\footnote{\url{https://www.swift.ac.uk/analysis/xrt/backscal.php}}, to correct for the 1-dimensional area ratio appropriate for WT mode data. With the command \texttt{quzcif}, we find the response matrix file, \texttt{swxwt0to2s6\_20110101v015.rmf}, in the \emph{Swift}/XRT CALDB. Ancillary response files are created with the tool \texttt{xrtmkarf} by using exposure maps (*.img) produced in the pipeline processing. All observations show count rates < 70 $\rm cts~s^{-1}$ (see the eighth column of Table \ref{table:1}) across 0.3-10.0 keV, indicating that no pile-up correction is necessary\footnote{\url{https://www.swift.ac.uk/analysis/xrt/pileup.php}}. All data are grouped to require at least 1 count per bin with command \texttt{ftgrouppha} in ``optmin'' mode (\citealt{Kaastra2016}). Spectra are fitted using \texttt{XSPEC v12.11.1} (\citealt{Arnaud1996}) with Cash-statistics (\citealt{Cash1979}), which is generally advised for Poisson-distributed data (\citealt{Humphrey2009}).

\subsection{Light-curves}
\label{section:2.2}
With \texttt{barycorr}\footnote{\url{https://www.swift.ac.uk/analysis/xrt/barycorr.php}}, we first perform a barycenter correction to the event data using the clock correction file \texttt{swclockcor20041120v148.fits}. The source and background regions are chosen in accordance with Section \ref{section:2.1}. We generate the light curves with a time bin size of 0.0036 s using the command \texttt{xselect} v2.4k\footnote{\url{https://www.swift.ac.uk/analysis/xrt/timing.php}}. \texttt{Stingray} v1.0 (\citealt{matteo2020}) is used to create power spectral density (PSD) data sets using all 0.3-10.0 keV events. For \emph{Swift/XRT}, we set the number of time-steps per interval to $2^{14}$ = 16,384, equivalent to 58.98 s. Accordingly, a periodogram with a minimum frequency of 0.017 Hz and a maximum (Nyquist) frequency of 139 Hz will be computed for each interval. The PSD data are logarithmically rebinned to 7\% frequency resolution to 94 bins each, as illustrated in Figure \ref{fig:4} (taking ObsId:00032461008 Seg1 as a representative example). The dead-time effect is also taken into account in Figure \ref{fig:4}. We present the PSD with Leahy normalization (\citealt{Leahy1983}), with the dead-time modified Poisson noise subtracted off (\citealt{Zhang1995}). The broadband RMS is computed from the 0.1-10.0 Hz integrated (\citealt{Miyamoto1991}, scaling the PSD appropriately by the count rate). The result of RMS is listed in the tenth column of Table \ref{table:1}. Dead time has been accounted for in all light curves.

\setcounter{figure}{4}
\begin{figure}
    \includegraphics[angle=0, width=\columnwidth]{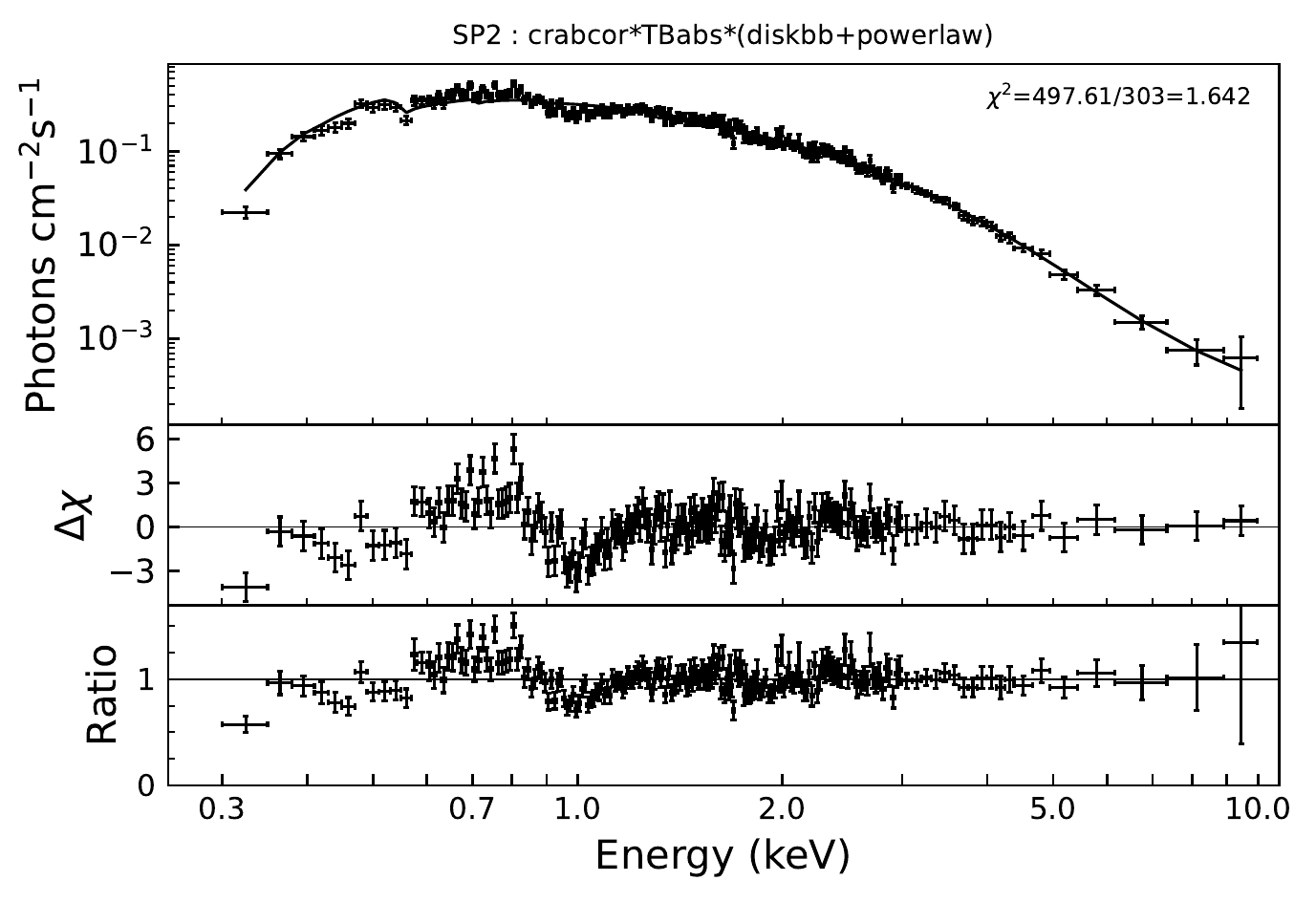}
    \caption{A spectral fit to representative dataset SP2, it demonstrates apparent emission and absorption features around 0.7 keV and 1.0 keV using model \texttt{crabcor*TBabs*(diskbb+powerlaw)}. For visual clarity, data has been rebinned. Top: spectral data and model. Middle: standardized fit residuals. Bottom: data-to-model ratio.}\label{fig:5}
\end{figure}

\setcounter{figure}{5}
\begin{figure}
\subfigure{
    \label{fig:6_1}
    \includegraphics[angle=0, width=\columnwidth]{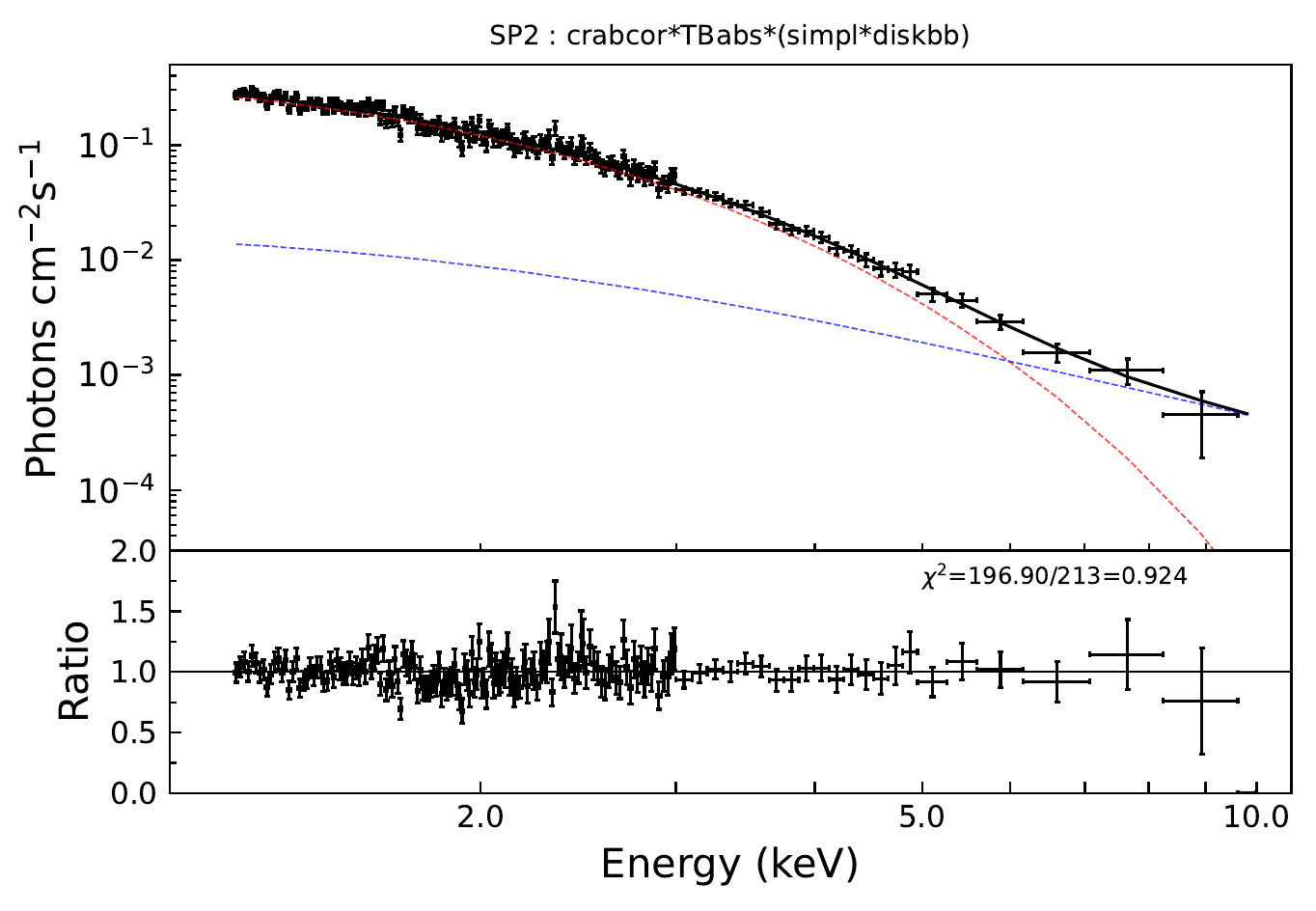}
    }
    \subfigure{
    \includegraphics[angle=0, width=\columnwidth]{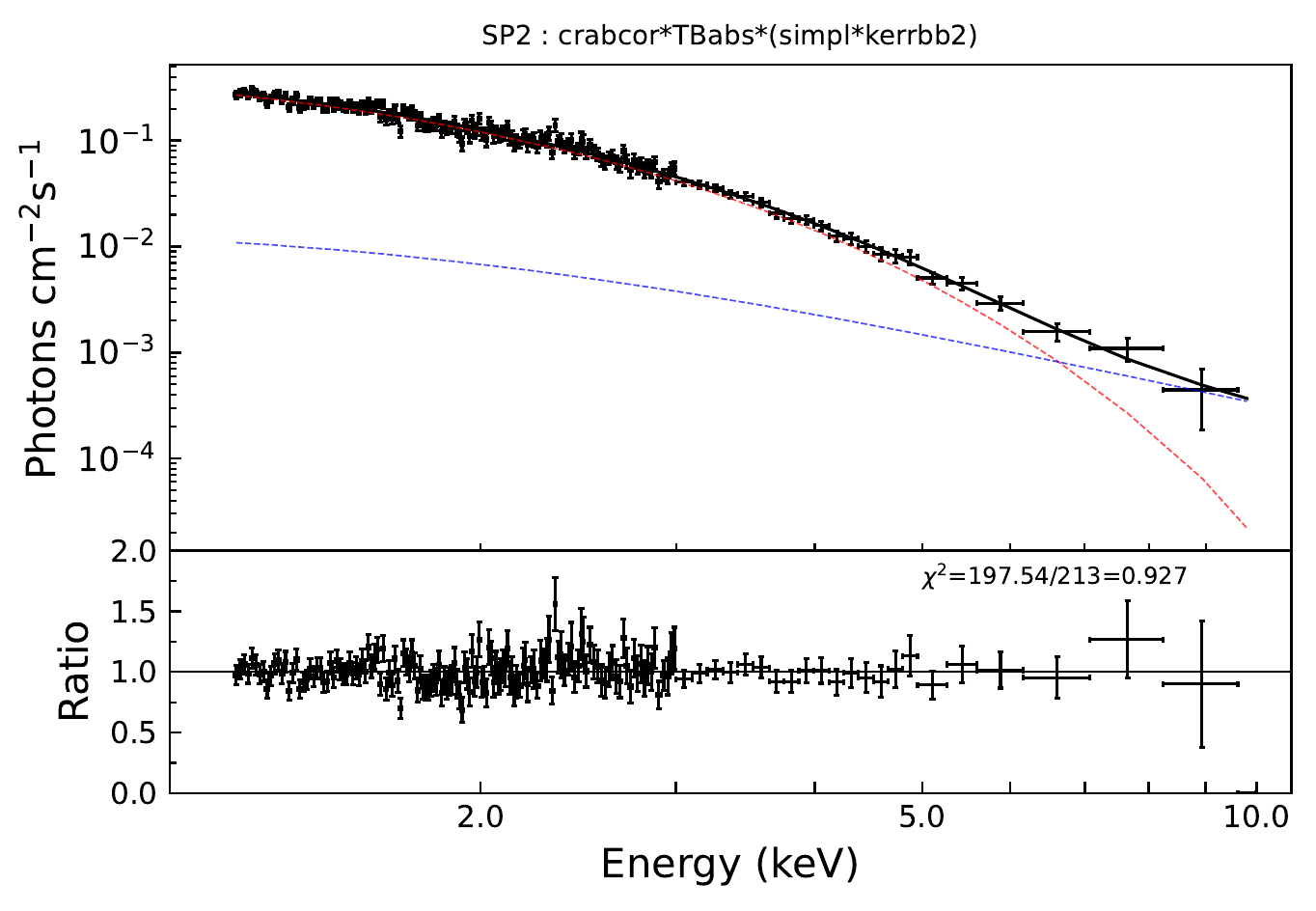}
      \label{fig:6_2}
    }
    \caption{Spectral fits for SP2 in 1.2-10.0 keV with model \texttt{crabcor*TBabs*(simpl*diskbb)} and model \texttt{crabcor*TBabs*(simpl*kerrbb2)}. The red dashed line represents the thermal component and the blue dashed line represents the Comptonization component. For visual clarity, data has been rebinned.}
\end{figure}


\setcounter{figure}{6}
\begin{figure}
\subfigure{
    \label{fig:7_1}
    \includegraphics[angle=0, width=\columnwidth]{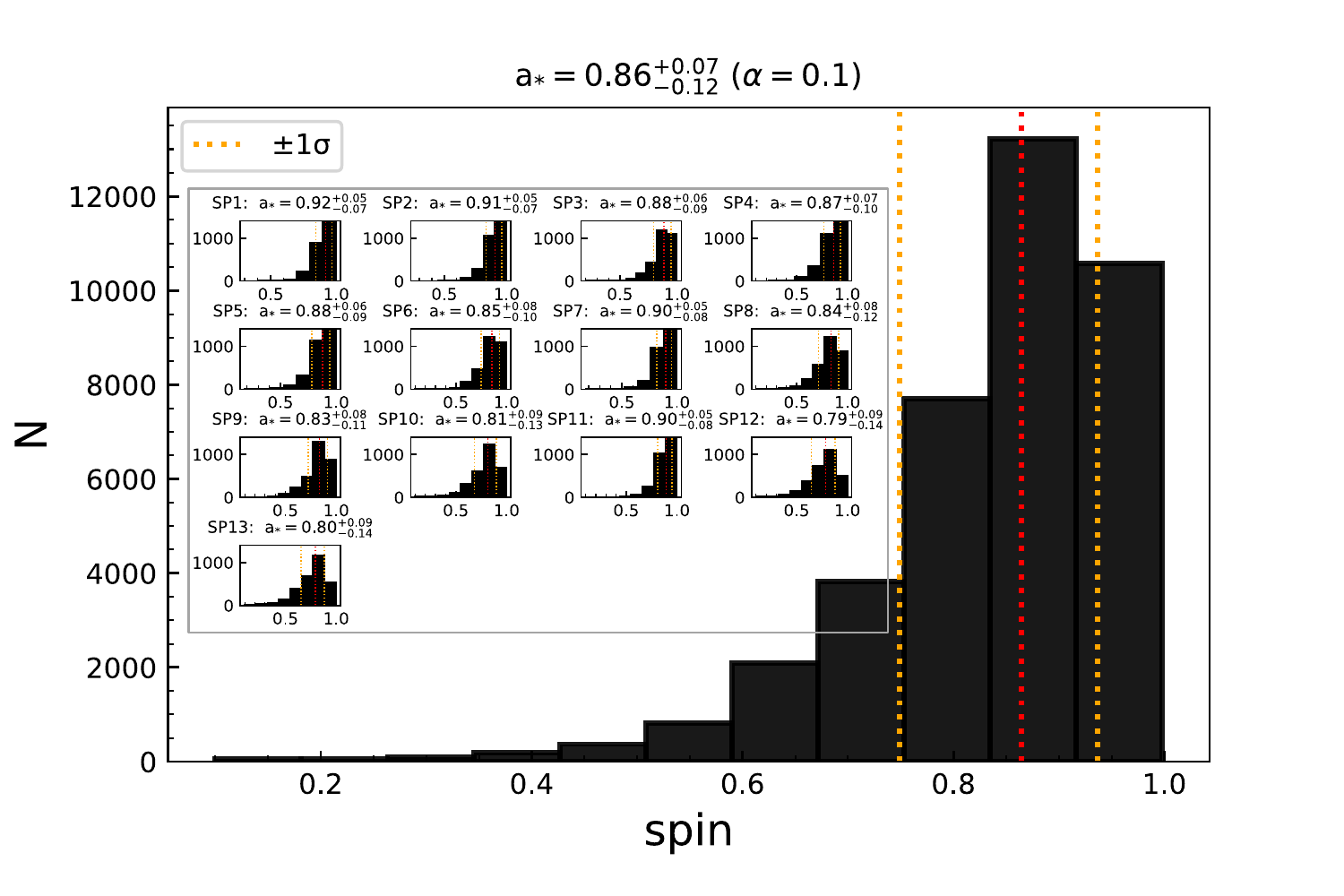}
    }
    \subfigure{
    \includegraphics[angle=0, width=\columnwidth]{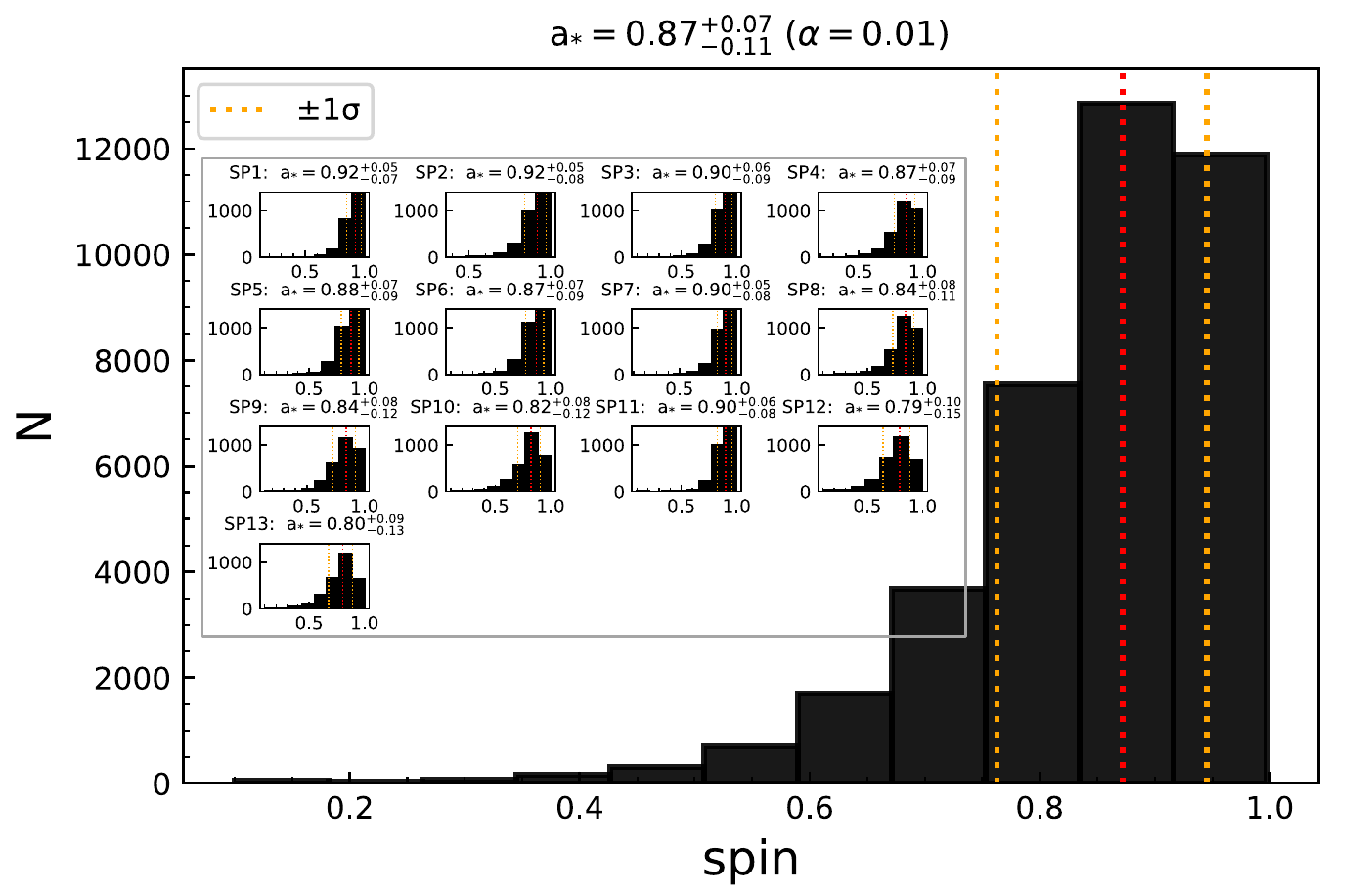}
      \label{fig:7_2}
      }
    \caption{a) For $\alpha=0.1$, the total histogram of $a_{*}$ for SP1-SP13, with 39,000 data points. The red dotted lines indicate the center value of $a_{*}$, while the orange dotted lines show the $\pm 68.3\% $ ($\pm$ 1$\sigma$). The insert of Figure \ref{fig:7_1} is spin results adopting $\alpha=0.1$, from SP1 to SP13, showing the results of the respective MC analyses. b) Same as Figure \ref{fig:7_1}, except for $\alpha= 0.01$.}
\end{figure}


\setcounter{figure}{7}
\begin{figure}
\subfigure{
    \label{fig:8_1}
    \includegraphics[angle=0, width=\columnwidth]{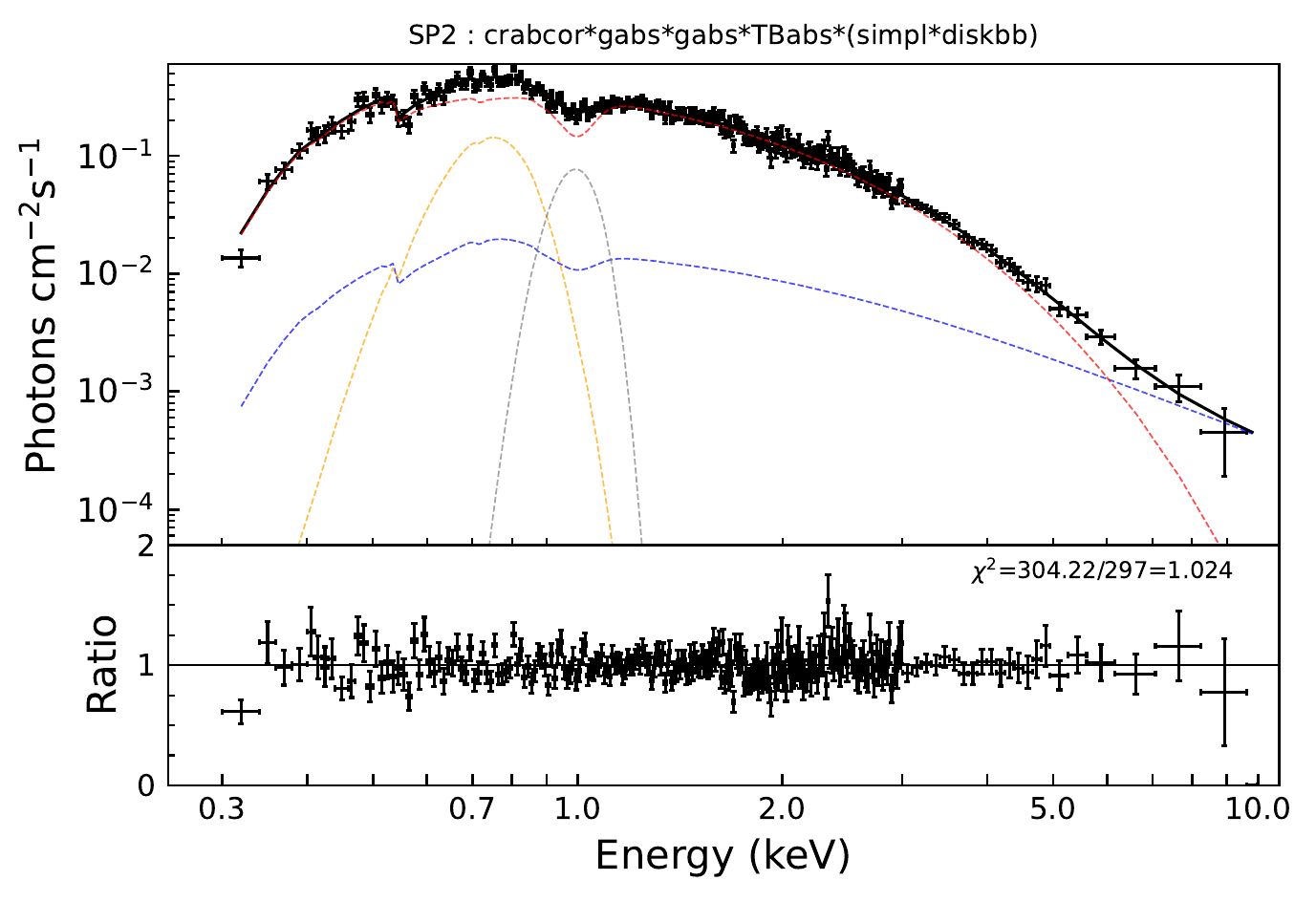}
    }
    \subfigure{
    \includegraphics[angle=0, width=\columnwidth]{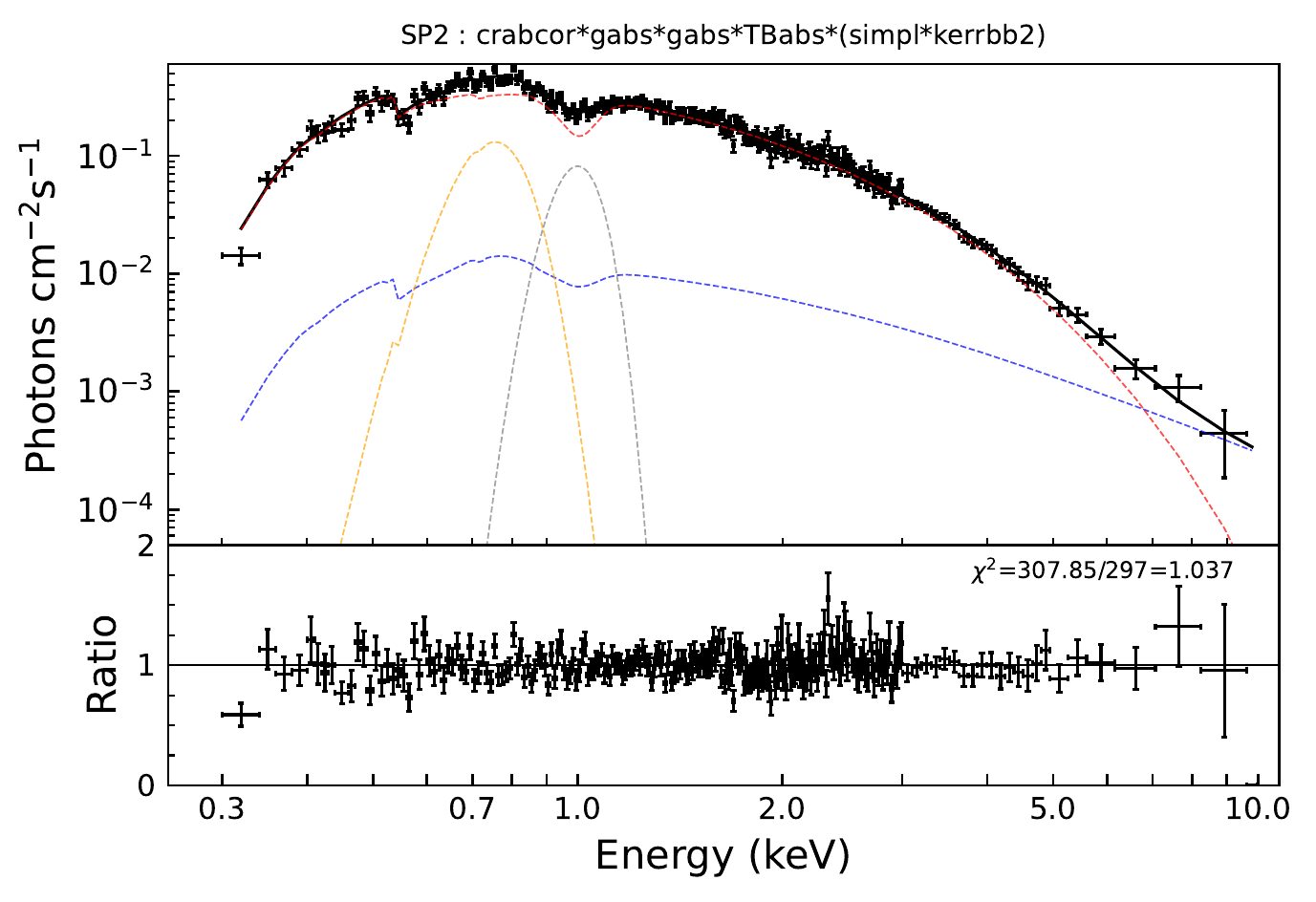}
      \label{fig:8_2}
    }
    \caption{Spectral fits for SP2 in 0.3-10.0 keV with model \texttt{crabcor*gabs*gabs*TBabs*(simpl*diskbb)} and model \texttt{crabcor*gabs*gabs*TBabs*(simpl*kerrbb2)}. The red dashed line represents the thermal component, the blue dashed line represents the Comptonization component, the yellow dotted line represents the emission line, and the gray dotted line represents the absorption line. For visual clarity, data has been rebinned.}
\end{figure}

\setcounter{figure}{8}
\begin{figure}
    \includegraphics[angle=0, width=\columnwidth]{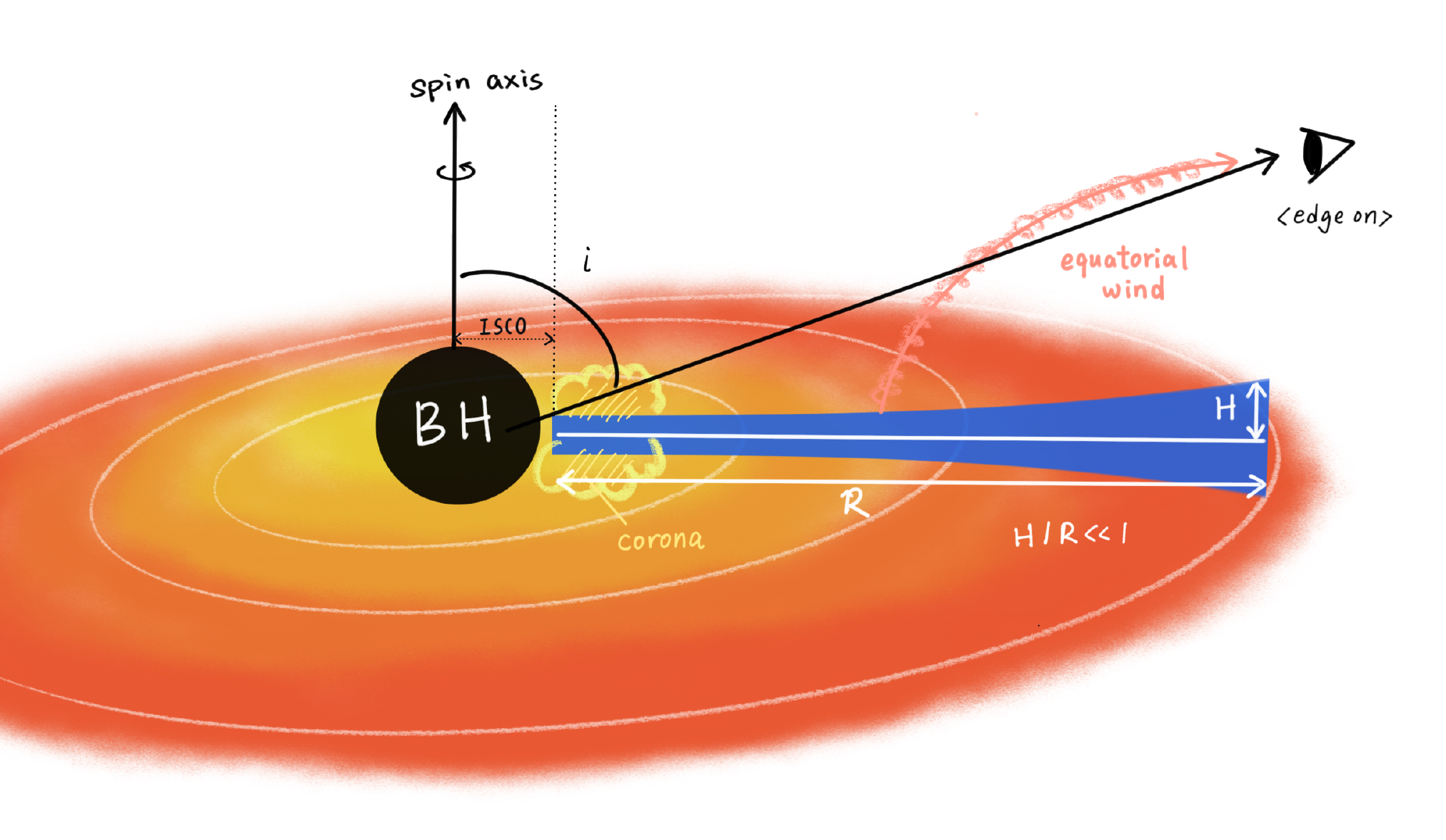}
    \caption{Equatorial wind illustration. Black represents the central black hole. Blue shows a schematic cross-section depicting the scale-height ratio H/R $\ll$ 1 for the geometrically-thin disk. Pink represents the equatorial wind. The central BH does likely illuminate the outer disk, and as a result, it may heat the outer disk, increasing the thermal pressure that forces away the wind that is flattened above the disk. Because of this, the wind is most readily detected for a high inclination case.}\label{fig:9}
\end{figure}

\setcounter{figure}{9}
\begin{figure}
\subfigure{
    \label{fig:10_1}
    \includegraphics[angle=0, width=\columnwidth]{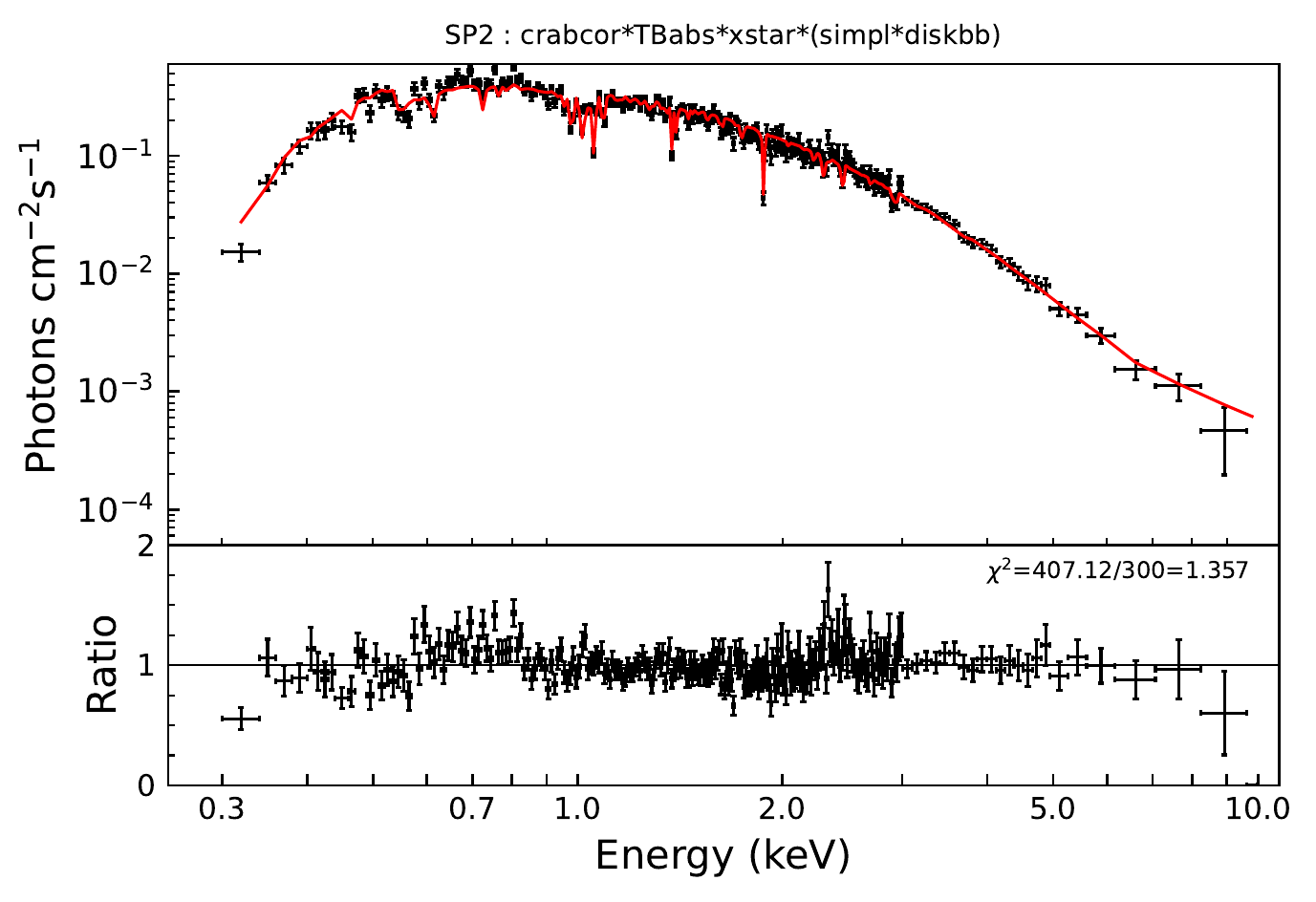}
    }
    \subfigure{
    \includegraphics[angle=0, width=\columnwidth]{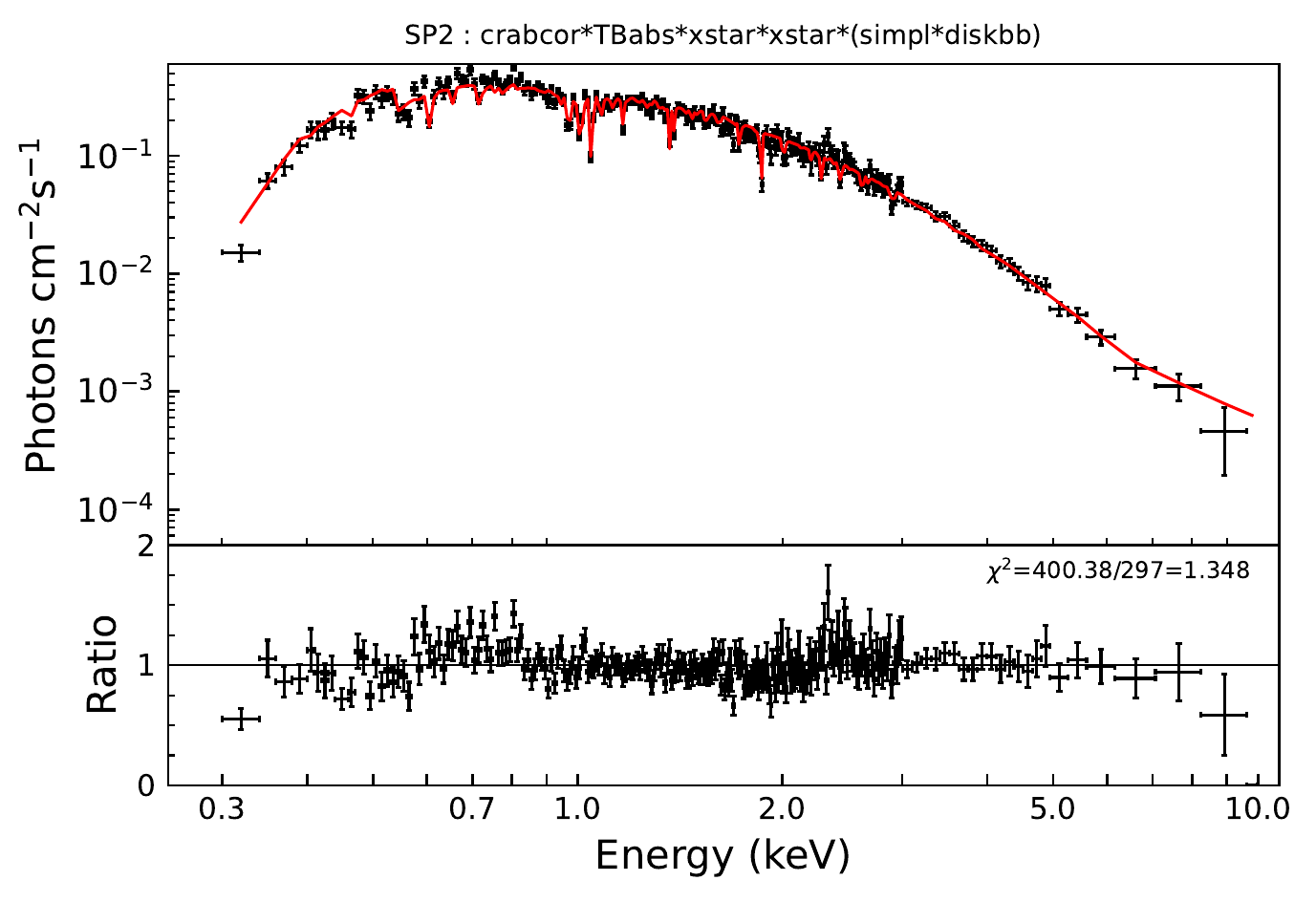}
      \label{fig:10_2}
    }
    \caption{Spectral fits for SP2 from 0.3-10.0 keV with model \texttt{crabcor*TBabs*xstar*(simpl*diskbb)} and model \texttt{crabcor*TBabs*xstar*xstar*(simpl*diskbb)}. For visual clarity, data has been rebinned.}
\end{figure}

\section{SPECTRAL ANALYSIS AND RESULTS}
\label{section:3}
Regardless of its selection or rejection by screening conditions, a preliminary and simplistic spectral model: \texttt{crabcor*TBabs*(diskbb+powerlaw)} is applied to all data to assess each spectrum in 0.3-10.0 keV, over which \emph{Swift}/XRT is sensitive. To account for the considerable flux-normalization discrepancies amongst X-ray missions, we follow prior work (\citealt{Steiner2010}) and standardize the calibration using the Crab as a reference. According to \url{https://www.swift.ac.uk/analysis/xrt/digest_cal.php#abs}, we find the parameter values of Crab for \emph{Swift}, and use $\Gamma=2.1$, $N=9.7 \text{ photons } s^{-1} \text{keV}^{-1}$ (\citealt{Toor1974}) as the standard values. Then we compute the normalization-correction coefficient $C_{\mathrm{TS}}$ is 0.948, and the slope difference $\Delta \Gamma_{\mathrm{TS}}$ is 0. The model \texttt{crabcor} is used to implement this standardization in \texttt{XSPEC}. We adopt \texttt{Tbabs} as the interstellar absorption model (for the interstellar medium, the photoionization cross-sections are based on \citealt{Verner1996}, and the abundances are based on \citealt{Wilms2000}). We fix the hydrogen column density ($N_{\mathrm H}$) to $1.8 \times 10^{21} \mathrm{cm^{-2}}$, which is the average of preliminary fits using free $N_{\mathrm H}$ for all data with model \texttt{crabcor*TBabs*(diskbb+powerlaw)}. It is also highly consistent with the total Galactic column density: \url{https://heasarc.gsfc.nasa.gov/cgi-bin/Tools/w3nh/w3nh.pl}. As seen in Figure \ref{fig:5}, there are two confounding emission/absorption components around 0.7 and 1.0 keV, respectively. In Section \ref{section:4.2} we examine potential wind models that explain these features and discuss their importance and impact on the spin. However, in our primary analysis, we here opt to omit this problematic energy range and perform our analysis over 1.2-10.0 keV. 

In summary, we perform a fit using the model \texttt{crabcor*TBabs*(simpl*diskbb)} for all stable data sets > 200s and RMS < 0.075, we use the resultant $f_{\mathrm{sc}}$ values as our final screening condition in 1.2-10.0 keV (see Section \ref{section:3.1} for details). We list $f_{\mathrm{sc}}$ only for a subset who meets the RMS <0.075 and the light-curve is stable in the last column of Table \ref{table:1}. We end up with a total of 13 ``gold" spectra that meet all these conditions. In addition, we also show the results of the fit for ``gold" spectra using the \texttt{powerlaw} model in Table \ref{table:2}.

\begin{table*}
    \renewcommand\arraystretch{1.3}
    \centering
	\caption{The best-fitting parameters with  \texttt{crabcor*TBabs*(diskbb+powerlaw)}}
    \begin{threeparttable}
    \begin{tabular}{ccccccccc}
    \hline
    \hline
Spec. & ObsID & Seg. &   \multicolumn{2}{c}{\texttt{diskbb}}& \texttt{powerlaw\tnote{1}}  & $\chi^2_\nu$ & $\chi^2$(d.o.f.)  & DF\\
   \cline{4-5}
 & & & $T_{\rm in}$ & $\rm norm$~$(\times 10^{2})$ &$\rm norm$~$(\times 10^{-2})$ & &&\\
  & & & ($\rm keV$) &  &($\rm photons ~\mathrm{keV^{-1}} \mathrm{cm^{-2}} \mathrm{s^{-1}}$ )& &&$(\%$)\\
\hline
SP1  &  00032339015 &  Seg2  &  0.862 $\pm$ 0.011  &  1.61  $\pm$ 0.07     &  13.31 $\pm$ 1.69     &  1.053  &  226.29/215 &  76.98    \\
SP2  &  00032461008 &  Seg1  &  0.774 $\pm$ 0.011  &  2.14  $\pm$ 0.10     &  6.56  $\pm$ 1.27     &  0.921  &  196.24/213 &  84.43    \\
SP3  &  00032461008 &  Seg2  &  0.755 $\pm$ 0.012  &  2.30  $\pm$ 0.13     &  9.20  $\pm$ 1.34     &  1.009  &  214.97/213 &  78.72    \\
SP4  &  00032339028 &  Seg1  &  0.729 $\pm$ 0.008  &  2.63  $\pm$ 0.10     &  10.21 $\pm$ 0.92     &  1.054  &  226.56/215 &  76.32    \\
SP5  &  00032461011 &  Seg2  &  0.722 $\pm$ 0.008  &  2.62  $\pm$ 0.11     &  7.92  $\pm$ 0.86     &  0.997  &  213.36/214 &  79.74    \\
SP6  &  00032461012 &  Seg1  &  0.718 $\pm$ 0.009  &  2.86  $\pm$ 0.12     &  9.03  $\pm$ 0.97     &  1.083  &  231.80/214 &  78.64    \\
SP7  &  00032339030 &  Seg1  &  0.709 $\pm$ 0.015  &  1.77  $\pm$ 0.14     &  11.65 $\pm$ 1.14     &  1.113  &  237.02/213 &  62.40    \\
SP8  &  00032461013 &  Seg1  &  0.749 $\pm$ 0.010  &  1.88  $\pm$ 0.09     &  22.40 $\pm$ 1.07     &  1.010  &  219.10/217 &  54.38    \\
SP9  &  00032461014 &  Seg1  &  0.727 $\pm$ 0.010  &  2.38  $\pm$ 0.11     &  17.49 $\pm$ 1.08     &  0.983  &  212.22/216 &  62.62    \\
SP10 &  00032461014 &  Seg2  &  0.725 $\pm$ 0.009  &  2.50  $\pm$ 0.11     &  19.92 $\pm$ 1.03     &  1.044  &  226.60/217 &  60.35    \\
SP11 &  00032461015 &  Seg1  &  0.705 $\pm$ 0.017  &  1.85  $\pm$ 0.16     &  11.14 $\pm$ 1.27     &  0.962  &  203.87/212 &  63.76    \\
SP12 &  00032461015 &  Seg3  &  0.738 $\pm$ 0.014  &  2.46  $\pm$ 0.16     &  26.32 $\pm$ 1.83     &  1.052  &  226.20/215 &  55.37    \\
SP13 &  00032461017 &  Seg1  &  0.701 $\pm$ 0.014  &  1.78  $\pm$ 0.12     &  25.29 $\pm$ 1.11     &  1.144  &  247.19/216 &  42.14    \\
               
    \hline
    \end{tabular}
    \begin{tablenotes}
        \footnotesize   
        \item{1} Gamma fixed at 2.2.
  \item Notes. In columns 4-9, we show the temperature of the inner disk radius ($T_{\rm in}$) in units of keV, the disk component's amplitude (norm), the normalization of power-law at 1.0 keV, the reduced chi-square ($\chi^2_\nu$), the total chi-square ($\chi^2$) and the degrees of freedom (d.o.f.), disk fraction (DF) gives the proportion of unabsorbed flux in the thermal component over the energy range from 1.2 to 10.0 keV. The errors are calculated in 68.3\% confidence intervals by \texttt{XSPEC}. 
    \end{tablenotes}
    \end{threeparttable}
    \label{table:2}
\end{table*}

\subsection{The non-relativistic spectral model}
\label{section:3.1}
All ``gold" spectra are fitted to a Comptonized multicolor disk blackbody (MCD) model with interstellar absorption. For the MCD, we use the \texttt{diskbb} model (\citealt{Mitsuda1984}; \citealt{Makishima1986}), and for Compton scattering, we use the \texttt{simpl} model (\citealt{Steiner2009}; \citealt{Steiner2009b}). We are unable to robustly constrain the photon index $\Gamma$ of the \texttt{simpl} model due to inadequate statistics in the hard X-ray band for much of the data. As a result, we fix $\Gamma$ to 2.2, which is a typical value for BHXRBs in the soft state (e.g., \citealt{Kolehmainen2011}). For the model \texttt{simpl}, we use the \texttt{energies} command in \texttt{XSPEC} to expand the sampled energies to 0.1-100.0 keV (see the appendix of \citealt{Steiner2009b}). The non-relativistic composite model \texttt{crabcor*TBabs*(simpl*diskbb)} is used. The best-fitting findings for SP1-SP13 are shown in Table \ref{table:3}. These 13 spectra are well fitted, and the reduced chi-square $\chi^2_\nu$ of SP1-SP13 are concentrated around 1. Figure \ref{fig:6_1} depicts the fitting of SP2 as a representation. We can easily see that the model fits well with no significant residuals.

\begin{table*}
    \renewcommand\arraystretch{1.3}
    \centering
	\caption{The best-fitting parameters with  \texttt{crabcor*TBabs*(simpl*diskbb)}}
    \begin{threeparttable}
    \begin{tabular}{ccccccccccc}
    \hline
    \hline
Spec. & ObsID & Seg. & \texttt{simpl}  &  \multicolumn{2}{c}{\texttt{diskbb}}& $\chi^2_\nu$ & $\chi^2$(d.o.f.) \\
   \cline{5-6} \cline{9-10}
 & & &  $f_{\rm sc}$ & $T_{\rm in} ~(\rm keV)$ & $\rm norm$~$(\times 10^{2})$ & & \\
\hline
SP1  &  00032339015 &  Seg2  &  0.098 $\pm$ 0.012  &  0.804 $\pm$ 0.014  &  2.51  $\pm$ 0.17     &  0.983  &  211.31/215 \\ 
SP2  &  00032461008 &  Seg1  &  0.051 $\pm$ 0.010  &  0.754 $\pm$ 0.013  &  2.62  $\pm$ 0.18     &  0.924  &  196.90/213 \\ 
SP3  &  00032461008 &  Seg2  &  0.072 $\pm$ 0.011  &  0.727 $\pm$ 0.014  &  3.08  $\pm$ 0.23     &  0.998  &  212.51/213 \\ 
SP4  &  00032339028 &  Seg1  &  0.076 $\pm$ 0.007  &  0.703 $\pm$ 0.009  &  3.57  $\pm$ 0.18     &  1.080  &  232.14/215 \\ 
SP5  &  00032461011 &  Seg2  &  0.064 $\pm$ 0.007  &  0.700 $\pm$ 0.009  &  3.37  $\pm$ 0.18     &  1.002  &  214.53/214 \\ 
SP6  &  00032461012 &  Seg1  &  0.067 $\pm$ 0.008  &  0.696 $\pm$ 0.010  &  3.72  $\pm$ 0.21     &  1.098  &  234.93/214 \\  
SP7  &  00032339030 &  Seg1  &  0.134 $\pm$ 0.014  &  0.660 $\pm$ 0.016  &  3.10  $\pm$ 0.30     &  1.105  &  235.38/213 \\ 
SP8  &  00032461013 &  Seg1  &  0.181 $\pm$ 0.009  &  0.669 $\pm$ 0.010  &  4.28  $\pm$ 0.26     &  0.939  &  203.80/217 \\  
SP9  &  00032461014 &  Seg1  &  0.135 $\pm$ 0.009  &  0.675 $\pm$ 0.010  &  4.21  $\pm$ 0.25     &  0.962  &  207.76/216 \\ 
SP10 &  00032461014 &  Seg2  &  0.147 $\pm$ 0.008  &  0.667 $\pm$ 0.009  &  4.70  $\pm$ 0.26     &  0.974  &  211.33/217 \\ 
SP11 &  00032461015 &  Seg1  &  0.126 $\pm$ 0.015  &  0.660 $\pm$ 0.018  &  3.11  $\pm$ 0.34     &  0.955  &  202.46/212 \\
SP12 &  00032461015 &  Seg3  &  0.171 $\pm$ 0.012  &  0.667 $\pm$ 0.014  &  5.29  $\pm$ 0.45     &  1.072  &  230.42/215 \\ 
SP13 &  00032461017 &  Seg1  &  0.239 $\pm$ 0.011  &  0.604 $\pm$ 0.013  &  5.49  $\pm$ 0.45     &  1.157  &  250.00/216 \\
               
    \hline
    \end{tabular}
    \begin{tablenotes}
        \footnotesize   
  \item Notes. Columns 4-9 show the scattered fraction ($f_{\rm sc}$), the temperature of the inner disk radius ($T_{\rm in}$) in units of keV, the disk component's amplitude (norm), the reduced chi-square ($\chi^2_\nu$), the total chi-square ($\chi^2$) and the degrees of freedom (d.o.f.). The errors are calculated in 68.3$\%$ confidence interval by \texttt{XSPEC}.  
    \end{tablenotes}
    \end{threeparttable}
    \label{table:3}
\end{table*}

\subsection{The relativistic spectral model}
\label{section:3.2}
Having now explored the data using preliminary non-relativistic models, we turn to the relativistic accretion disk model \texttt{kerrbb2} (\citealt{McClintock2006}). \texttt{kerrbb2} is a combination of the disk models \texttt{bhspec}\footnote{\url{https://www.cita.utoronto.ca/~swd/xspec.html}} and \texttt{kerrbb}. \texttt{bhspec} is used to derive the spectral hardening factor $f \equiv T \mathrm{col / T \mathrm{eff}}$ (\citealt{Davis2005}), whilst \texttt{kerrbb} is used to model the disk using ray-tracing computations (\citealt{Li2005}). We begin by generating spectral-hardening look-up tables using \texttt{bhspec} with a default viscosity parameter value $\alpha=0.1$\footnote{the only available values for viscosity parameters are 0.1 and 0.01.} as a conservative estimate. The $f$-table is read in and used to automatically set the value of $f$ during \texttt{kerrbb2} as a function of $\dot{M}$ and $a_{*}$. The complete model can be expressed as \texttt{crabcor*TBabs*(simpl*kerrbb2)}. For model \texttt{kerrbb2}, the self-irradiation of the disk (rflag=1) and the influence of limb-darkening (lflag=1) are taken into account. And we also set the torque at the inner boundary of the disk to zero (eta=0). Table \ref{table:4} summarizes our results, which implies that a moderate spin is most plausible. We can also see from the findings of $L/ L_{\mathrm{Edd}}$ in Table \ref{table:4} that it is in accordance with our demands of the disk being geometrically thin by requiring that the dimensionless luminosity < ~0.3 (\citealt{McClintock2014}).


\begin{table*}
    \renewcommand\arraystretch{1.3}
    \centering
	\caption{The best-fitting parameters with the relativistic model \texttt{crabcor*TBabs*(simpl*kerrbb2)}}
    \begin{threeparttable}
    \begin{tabular}{cccccccccc}
    \hline
    \hline
     viscosity  & Spec. & ObsID & Seg. & \texttt{simpl}  &  \multicolumn{2}{c}{\texttt{kerrbb2}}& $\chi^2_\nu$ & $\chi^2$(d.o.f.) & $L/ L_{\mathrm{Edd}}$ \\
   \cline{6-7}
$\alpha$ &  & & & $f_{\rm sc}$ & $a_{*}$ & $\dot{M} ~(\times 10^{18} \mathrm{~g} \mathrm{~s}^{-1})$ & & &\\
\hline
&SP1  &  00032339015 &  Seg2  &  0.079  $\pm$ 0.012     &  0.917  $\pm$ 0.009     &  0.138  $\pm$ 0.006     &  1.037     &  222.87/215  &  0.018 \\    
&SP2  &  00032461008 &  Seg1  &  0.037  $\pm$ 0.010     &  0.909  $\pm$ 0.009     &  0.117  $\pm$ 0.005     &  0.927     &  197.53/213  &  0.015 \\    
&SP3  &  00032461008 &  Seg2  &  0.059  $\pm$ 0.011     &  0.884  $\pm$ 0.011     &  0.131  $\pm$ 0.006     &  1.023     &  217.87/213  &  0.015  \\   
&SP4  &  00032339028 &  Seg1  &  0.063  $\pm$ 0.007     &  0.863  $\pm$ 0.010     &  0.143  $\pm$ 0.005     &  1.065     &  228.98/215  &  0.016   \\  
&SP5  &  00032461011 &  Seg2  &  0.052  $\pm$ 0.007     &  0.875  $\pm$ 0.008     &  0.128  $\pm$ 0.004     &  1.004     &  214.82/214  &  0.014     \\
&SP6  &  00032461012 &  Seg1  &  0.054  $\pm$ 0.007     &  0.856  $\pm$ 0.010     &  0.147  $\pm$ 0.005     &  1.090     &  233.18/214  &  0.016     \\
0.1&SP7  &  00032339030 &  Seg1  &  0.120  $\pm$ 0.013     &  0.899  $\pm$ 0.014     &  0.084  $\pm$ 0.005     &  1.109     &  236.18/213  &  0.010  \\   
&SP8  &  00032461013 &  Seg1  &  0.164  $\pm$ 0.010     &  0.839  $\pm$ 0.010     &  0.153  $\pm$ 0.005     &  0.964     &  209.08/217  &  0.016 \\    
&SP9  &  00032461014 &  Seg1  &  0.120  $\pm$ 0.009     &  0.836  $\pm$ 0.010     &  0.157  $\pm$ 0.005     &  0.979     &  211.41/216  &  0.016 \\    
&SP10 &  00032461014 &  Seg2  &  0.130  $\pm$ 0.007     &  0.817  $\pm$ 0.013     &  0.176  $\pm$ 0.006     &  1.013     &  219.91/217  &  0.017 \\    
&SP11 &  00032461015 &  Seg1  &  0.113  $\pm$ 0.014     &  0.896  $\pm$ 0.017     &  0.086  $\pm$ 0.006     &  0.960     &  203.56/212  &  0.010  \\   
&SP12 &  00032461015 &  Seg3  &  0.151  $\pm$ 0.013     &  0.787  $\pm$ 0.018     &  0.214  $\pm$ 0.010     &  1.057     &  227.25/215  &  0.020  \\   
&SP13 &  00032461017 &  Seg1  &  0.219  $\pm$ 0.011     &  0.797  $\pm$ 0.017     &  0.145  $\pm$ 0.007     &  1.139     &  246.12/216  &  0.014 \\
\hline
&SP1  &  00032339015 &  Seg2  &  0.080  $\pm$ 0.012     &  0.924  $\pm$ 0.008     &  0.134  $\pm$ 0.006     &  1.035     &  222.47/215  &  0.018\\     
&SP2  &  00032461008 &  Seg1  &  0.037  $\pm$ 0.010     &  0.917  $\pm$ 0.009     &  0.112  $\pm$ 0.005     &  0.927     &  197.50/213  &  0.014     \\
&SP3  &  00032461008 &  Seg2  &  0.058  $\pm$ 0.010     &  0.896  $\pm$ 0.012     &  0.126  $\pm$ 0.006     &  1.022     &  217.68/213  &  0.015 \\    
&SP4  &  00032339028 &  Seg1  &  0.064  $\pm$ 0.007     &  0.873  $\pm$ 0.008     &  0.139  $\pm$ 0.004     &  1.064     &  228.75/215  &  0.016  \\   
&SP5  &  00032461011 &  Seg2  &  0.052  $\pm$ 0.007     &  0.882  $\pm$ 0.008     &  0.125  $\pm$ 0.004     &  1.004     &  214.86/214  &  0.014 \\    
&SP6  &  00032461012 &  Seg1  &  0.054  $\pm$ 0.007     &  0.863  $\pm$ 0.011     &  0.144  $\pm$ 0.006     &  1.089     &  233.09/214  &  0.016 \\    
0.01 &SP7  &  00032339030 &  Seg1  &  0.120  $\pm$ 0.014     &  0.907  $\pm$ 0.014     &  0.082  $\pm$ 0.005     &  1.108     &  236.11/213  &  0.010 \\    
&SP8  &  00032461013 &  Seg1  &  0.163  $\pm$ 0.009     &  0.844  $\pm$ 0.010     &  0.150  $\pm$ 0.005     &  0.964     &  209.12/217  &  0.016 \\    
&SP9  &  00032461014 &  Seg1  &  0.120  $\pm$ 0.009     &  0.842  $\pm$ 0.010     &  0.154  $\pm$ 0.005     &  0.979     &  211.48/216  &  0.016  \\   
&SP10 &  00032461014 &  Seg2  &  0.130  $\pm$ 0.008     &  0.825  $\pm$ 0.013     &  0.172  $\pm$ 0.006     &  1.012     &  219.65/217  &  0.017 \\    
&SP11 &  00032461015 &  Seg1  &  0.114  $\pm$ 0.015     &  0.905  $\pm$ 0.016     &  0.083  $\pm$ 0.006     &  0.960     &  203.45/212  &  0.010 \\    
&SP12 &  00032461015 &  Seg3  &  0.151  $\pm$ 0.013     &  0.795  $\pm$ 0.018     &  0.210  $\pm$ 0.010     &  1.057     &  227.24/215  &  0.020 \\    
&SP13 &  00032461017 &  Seg1  &  0.220  $\pm$ 0.011     &  0.805  $\pm$ 0.018     &  0.142  $\pm$ 0.007     &  1.140     &  246.15/216  &  0.014 \\                 
    \hline
    \end{tabular}
    \begin{tablenotes}
        \footnotesize   
  \item Notes. Columns 5-10 show the scattered fraction ($f_{\rm sc}$), the dimensionless spin parameter ($a_{*}$), the mass accretion rate through the disk in units of 10$^{18}$ g s $^{-1}$ ($\dot{M}$), the reduced chi-square ($\chi_{\nu}^{2}$), the total chi-square ($\chi^2$) and the degrees of freedom (d.o.f.), the bolometric Eddington-scaled luminosity ($L/ L_{\mathrm{Edd}}$, where $L_{\mathrm{Edd}}= 1.3 \times 10^{38}\left(M / M_{\odot}\right)$ erg $\mathrm{s}^{-1}$). The errors are calculated in 68.3$\%$ confidence interval by \texttt{XSPEC}. For these fits, the system parameters are $M_{\rm BH}=8.9_{-1.0}^{+1.6} M_{\odot}$, $D=7.5_{-1.4}^{+1.8} \mathrm{kpc}$, and $i={72_{-8}^{+5}}^{\circ}$ (uncertainties are 1 $\sigma$; \citealt{Mata2021}).
    \end{tablenotes}
    \end{threeparttable}
    \label{table:4}
\end{table*}

\subsection{Comprehensive spin error analysis}
\label{section:3.3}
In this section, we consider various sources of observational error, both systematic and statistical (See Section 5 and Appendix A in \citealt{Steiner2011}; Section 5 in \citealt{McClintock2014} for details), that have an impact on our final estimate of the spin. They are (1) the uncertainties from the analytic Novikov-Thorne model and the disk atmosphere model; For MAXI J1305-704, these model errors are especially small because of the low luminosity of the disk, and so we do not explore its effect in this paper. (2) the effect of the emission/absorption lines in low-energy bands (discussion see Section \ref{section:4.1}); (3) the effect of hydrogen column density $N_{\mathrm{H}}$ (discussion see Section \ref{section:4.2}); (4) the influence of the fixed photon index $\Gamma$ (discussion see Section \ref{section:4.3}); (5) the influence of viscosity parameter $\alpha $ (discussion see Section \ref{section:4.4}); (6) X-ray flux calibration uncertainties; In general, we need to include the uncertainty in the luminosity due to the $\sim$10\% uncertainty in the flux of the Crab (\citealt{Toor1974}). But for MAXI J1305-704, the effect of the uncertainty in the absolute flux calibration is much smaller in impact compared to the 24\% uncertainty in $D$, which is equivalent to a 48\% uncertainty in the measurement of flux. Therefore, absolute X-ray flux calibration uncertainties are of minor importance compared to the uncertainties in the input system parameters. and (7) the uncertainties in the input parameters $M$, $i$, and $D$. Specifically, as in earlier works (\citealt{Liu2008}; \citealt{Gou2009}), we employ Monte Carlo (MC) approach to assess the uncertainties in the CF method, from uncertainty in the dynamical inputs ($M$, $i$, $D$). Due to the measurement correlations between $M$ and $i$, the mass function $f(M)=M^{3} \sin ^{3} i /\left(M_{\mathrm{opt}}+M\right)^{2}$ (where $M$ is the mass of the black hole, $M_{\mathrm{opt}}$ is the mass of the optical companion, and $i$ is the accretion disk inclination) is used to draw random parameters, as described in \citealt{Gou2011}. Specifically, we adopt $M_{\mathrm{opt}}=0.43 \pm 0.16 M_{\odot}$ (\citealt{Mata2021}) to generate 3000 sets of ($M_{\mathrm{opt}}$, $f(M)$, $i$, $D$) for SP1-SP13, assuming each parameter is independent and has a Gaussian distribution. For each ($M_{\mathrm{opt}}$, $f(M)$, $i$), we determine $M$ via the mass function so that the parameters in these sets become ($M$, $i$, $D$). Then, for each set of parameters, we calculate $f$ via look-up tables with the viscosity parameter $\alpha$ fixed at 0.1 and 0.01 separately. Finally, we fit 3000 data sets to the composite model \texttt{crabcor*TBabs*(simpl*kerrbb2)} (see Section \ref{section:3.2}) to obtain the spin distribution and quantify the errors. The MC-determined error analysis of each spectrum is depicted in the inset of Figure \ref{fig:7_1} ($\alpha $=0.1) and \ref{fig:7_2} ($\alpha $=0.01). The histograms of $a_{*}$ for SP1-SP13 are depicted in Figure \ref{fig:7_1} ($\alpha $=0.1) and \ref{fig:7_2} ($\alpha $=0.01). Values and their 1 $\sigma$ equivalent uncertainties depict the mode and 68\% confidence interval centered on the mode. [median is the value in the middle, and the mode is the high point of a Gaussian distribution.] Consequently, we perform a comprehensive analysis in which we find that the spin of MAXI J1305-704 has an acceptable range of 0.74-0.95 (1 sigma confidence interval). We find that the individual spectra exhibit surprising variance in their spin determination (compared to other sources); many are centered on moderate values whereas others seem to favor the maximum spin. Since the 13 spectra we selected are stable, such differences are not due to variability. Possible reasons are discussed in detail in Section \ref{section:4.5}.

\section{DISCUSSIONS}
\label{section:4}
\subsection{The influence of the low energy band}
\label{section:4.1}

\subsubsection{Adding simplistic gaussian models: \texttt{gabs}}
\label{section:4.1.1}
Here we examine the impact of the low-energy features on our fitting results by alternatively considering empirical and physical models for their origin. We begin using a simplistic Gaussian model (\texttt{gabs}) to capture the two low-energy features, and by allowing both positive and negative normalizations allow for features to be either emission or absorption. For 0.3-10.0 keV, the complete non-relativistic model is \texttt{crabcor*gabs*gabs*TBabs*(simpl*diskbb)}, and the complete relativistic model is \texttt{crabcor*gabs*gabs*TBabs*(simpl*kerrbb2)}. To facilitate comparison, we take SP2 as our touchstone example, which is representative of the full ``gold" data sets. We can see from Figure \ref{fig:8_1} and \ref{fig:8_2} that the data fits are of similar quality and character using both non-relativistic and relativistic models. The best-fit results are listed in Table \ref{table:5}. From the fourth and fifth columns of Table \ref{table:5} (Models A and B, respectively), we can see that when fitting to different energy bands, $f_{\rm sc}$ and $T_{\rm in}$ remain the same. Similarly, columns 6 and 7 show that with Models C and D, we can see that the spin changes by 3\%. These results suggest that ignoring the lowest energies has a minor impact on the fit results, and is a reasonable simplification.

\begin{table*}\small
    \renewcommand\arraystretch{1.5}
    \caption{The best-fitting parameters for SP2 in both energy bands}
	
    \begin{threeparttable}
    \resizebox{\textwidth}{!}{
    \begin{tabular}{ccccccccc}
    
    \hline
    \hline
Component  & Parameter & Description & A & B & C & D & E & F \\
 & & & 1.2-10.0 keV & 0.3-10.0 keV & 1.2-10.0 keV ($\alpha=0.1$) & 0.3-10.0 keV ($\alpha=0.1$) & 0.3-10.0 keV & 0.3-10.0 keV\\
\hline
\tt{gabs1}      & $E_{\text {line }}(\mathrm{keV}) $                                             &line energy                                                       &	...                                            &0.74 $\pm$0.02                       &	...                                     &0.75$\pm$0.02               &...                                 &...             \\
		    & $\sigma_{\text {line }}(\times 10^{-2} ~\mathrm{keV}) $            &line width                                                          &	...                                            &9.7 $\pm$2.9                    &	...                                     &8.0 $\pm$2.8          &...                                 &...      \\
		    & $\tau ~(\times 10^{-2})$                                                            &line depth                                                          &	...                                            &-8.9 $\pm$2.1                  &	...                                     &-6.5 $\pm$1.6        &...                                  &...         \\
\tt{gabs2}      & $E_{\text {line }}(\mathrm{keV})$                                            &line energy                                                         &	...                                            &1.00 $\pm$0.03                       &	...                                     &1.00 $\pm$0.02              &...                                  &...           \\
		    & $\sigma_{\text {line }}(\times 10^{-2} ~\mathrm{keV}) $           &line width                                                           &	...                                            &6.4 $\pm$2.7                   &	...                                      &6.8 $\pm$2.4         &...                                 &...        \\
		    & $\tau ~(\times 10^{-2})$                                                           & line depth                                                          &	...                                            &4.5 $\pm$1.7                   &	...                                      &5.0 $\pm$1.4         &...                                 &...       \\
\tt{xstar1}     & $N(\times 10^{22} \mathrm{~cm}^{-2})$                                    &column density                                                 &	...                                            &...                                             &	...                                      &...                                   &6.0$\pm$2.9        &4.1$\pm$0.9      \\
		    & $\log \xi(\times \mathrm{erg}\mathrm{~cm}\mathrm{~s}^{-1}/s)$ &ionization                                                      &	...                                            &...                                             &	...                                       &...                                  &2.61$\pm$0.20            & 2.49$\pm$0.11 \\
		    &z $(\times 10^{-2})$                                                                    & gravitational red-shift                                      &	...                                            &...                                              &	...                                       &...                                  &6.9$\pm$0.8        & 7.8$\pm$1.2       \\		    
\tt{xstar2}      & $N(\times 10^{22} \mathrm{~cm}^{-2})$                                  &column density                                                  &	...                                            &...                                             &	...                                        &...                                 &...                                 &$13.74~(*)$       \\
		    & $\log \xi$$(\mathrm{~erg} \mathrm{~cm} \mathrm{~s}^{-1})$  &ionization                                                   &	...                                            &...                                             &	...                                       &...                                 &...                                 & 3.68$\pm$1.51  \\
		    & z                                                                                                & gravitational red-shift                                       &	...                                            &...                                             &	...                                        &...                                 &...                                &      0$\pm$0.03   \\	    		    		    		    
\tt{simpl}       & $f_{\rm sc}$                                                                              &the scattered fraction                                         &    0.051 $\pm$ 0.010               &0.049$\pm$0.010                   &0.037  $\pm$ 0.010             &0.034$\pm$ 0.010       &0.062$\pm$0.009       &    0.067$\pm$0.011         \\
\tt{diskbb}     & $T_{\rm in}~(\mathrm{keV})$                                                   &the temperature of the inner disk radius             &     0.754 $\pm$ 0.013              &0.757$\pm$0.013                  &         ...                               &            ...                       &0.705$\pm$0.010      &   0.699$\pm$0.011       \\
		    & $\rm norm$~$(\times 10^{2})$                                                 &normalization                                                       &     2.62  $\pm$ 0.18                &  2.59  $\pm$ 0.17                  &     ...                                  &     ...                              &4.06$\pm$0.22          & 4.42$\pm$0.87              \\
\tt{kerrbb2}   & $a_{*}$                                                                                     &the dimensionless spin parameter                      &     ...                                        &...                                             & 0.909  $\pm$ 0.009         &0.936$\pm$0.007          &...                                &...           \\
		    &  $\dot{M} ~(\times 10^{18} \mathrm{~g} \mathrm{~s}^{-1})$   &the effective mass accretion rate of the disk       &     ...                                        & ...                                            &  0.117  $\pm$ 0.005        &0.101$\pm$0.004          &...                                 &...         \\
\hline
$L/ L_{\mathrm{Edd}}$            &...                                                                     &  the bolometric Eddington-scaled luminosities    &...	                                      &...	                                      &  0.015                   	         & 0.014                          &...                                &...       \\
$\chi_{v}^{2}$                          &...                                                                     &  ...                                                                        & 0.924                        	      &1.024	                               & 0.927                    	          &1.037                          &1.357                          &1.348 \\
$\chi^{2} ~(\mathrm{d.o.f.})$   &  ...                                                                    &  ...                                                                        &196.90/213                           &304.22/297                            &197.54/213                          &307.85/297                  &407.12/300                  &    400.38/297   \\
    \hline      
    \end{tabular}}
    \begin{tablenotes}
        \footnotesize 
  \item Notes. Model A is \texttt{crabcor*TBabs*(simpl*diskbb)} in 1.2-10.0 keV, while Model B is \texttt{crabcor*gabs*gabs*TBabs*(simpl*diskbb)} in 0.3-10.0 \\keV. Model C is \texttt{crabcor*TBabs*(simpl*kerrbb2)} in 1.2-10.0 keV, and Model D is \texttt{crabcor*gabs*gabs*TBabs*(simpl*kerrbb2)} in 0.3-10.0 \\keV. Model E is \texttt{crabcor*TBabs*xstar*(simpl*diskbb)} in 0.3-10.0  keV, and Model F is \texttt{crabcor*TBabs*xstar*xstar*(simpl*diskbb)} in \\0.3-10.0 keV. (*) indicates that the parameter is unconstrained; the central value of the fit is provided for reference. \\The errors are calculated in 68.3$\%$ confidence interval by \texttt{XSPEC}.  
    \end{tablenotes}
    \end{threeparttable}
    \label{table:5}
\end{table*}

\subsubsection{Considering the influence of equatorial wind}\label{section:4.1.2}
We note that the broadened low-energy features can not be explained by the instrumental origins, so we now examine more physically-rigorous models for the low-energy features, especially consideration of a wind origin. Equatorial winds in BHXRBs are commonly indicated by the presence of highly-ionized absorption lines. As in the case of the broad emission/absorption can plausibly be interpreted as relativistic disk lines (0.7 keV and 1.0 keV lines may be relativistic O VIII and Fe L lines respectively) due to the equatorial wind, similar winds were launched previously like in H1743-322 (\citealt{miller2006}) and GRS 1915+105 (\citealt{ueda2009}). This phenomenon is often found in high-inclination (also called edge-on view, as shown in Figure 1 of \citealt{Moriyama2017}) systems in the soft state (for more detailed information, see \citealt{Ponti2012}). We show a cartoon of the wind on the equatorial plane (see Figure \ref{fig:9}), which can help to illustrate this process. 

We also note that for MAXI J1305-704, broad emission/absorption line features are also observed at the low energy end in the \emph{Chandra} gratings data (\citealt{Miller2014}), \emph{Suzaku} (\citealt{Shidatsu2013}) and \emph{Swift} (different from the ``gold" spectra used in our text) observations. \citealt{Miller2014} suggested that these features could be attributed to a failed disk wind which was assessed using a large \texttt{XSTAR} grid. We follow \citealt{Miller2014}'s analysis and construct similar (though lower-resolution) \texttt{XSTAR} grids, using the same settings. Specifically, we assume a simple unabsorbed incident spectrum of kT = 1.0 keV with a luminosity of $L=1.0 \times 10^{37} \operatorname{erg~s}^{-1}$ in the energy range of 1-1000 Ry (1 Ry=13.6 eV). We use \texttt{XSTAR} v2.58e (\citealt{Kallman2001}) and assume a covering factor of $\Omega / 4 \pi=0.5$, turbulent velocity of $v=700 \mathrm{~km} / \mathrm{s}$, and all elements of solar abundances. Then we construct a grid of 150 models, spanning $2.0 \leq \log \xi $~(ionization)~$ \leq 4.5$, $5 \times 10^{21} \mathrm{~cm}^{-2} \leq N$~(column density)~$\leq 5 \times 10^{23} \mathrm{~cm}^{-2}$, and $10^{16} \mathrm{~cm}^{-3} \leq n$~(density)~$ \leq 10^{18} \mathrm{~cm}^{-3}$. The fit is much improved by using \texttt{crabcor*TBabs*xstar*(simpl*diskbb)} in 0.3-10.0 keV, when assuming $ n=10^{17} \mathrm{~cm}^{-3}$ (\citealt{Miller2014}). However, it is still far from being formally accepted as a statistical fit. The results of the fit are listed in the eighth column of Table \ref{table:5}. From Figure \ref{fig:10_1}, we can clearly see that the residuals around 0.7 keV are not eliminated. We explore this deficiency by allowing for an additional zone by adding an additional \texttt{XSTAR} model. The full model is \texttt{crabcor*TBabs*xstar*xstar*(simpl*diskbb)}. And also assuming $ n=10^{17} \mathrm{~cm}^{-3}$. The fitting results are presented in Table \ref{table:5} and Figure \ref{fig:10_2}. The use of two \texttt{XSTAR} models did not significantly improve the 0.7 keV residuals compared to the use of only one. We conclude that the residuals at 0.7 and 1.0 keV are not well-fitted concurrently for two winds with equal density. Addressing this deficiency may be possible by using two \texttt{XSTAR} models with a large difference in density. However, such a detailed analysis of these features is outside the scope of our work. We note that where the present \texttt{XSTAR} models do not perfectly fit the spectrum, we still consider the influence of the ionized absorber(s) on the spin. We find that, taking SP2 as a representative example, the spin increases by $\Delta a_* = 0.024$ with model \texttt{crabcor*TBabs*xstar*xstar*(simpl*kerrbb2)} ($\chi_{v}^{2}$=439.59/297=1.480) in $\alpha=0.1$ case.


\subsection{The influence of hydrogen column density $N_{\mathrm{H}}$}
\label{section:4.2}
To check if the hydrogen column density has any influence on the spin results, we explore a generous range of putative values for $N_{\mathrm{H}}$, such as 0.10, 0.30, and 0.40 in units of $10^{22} \mathrm{cm}^{-2}$. Table \ref{table:6} shows the detailed fitting results. When $N_{\mathrm{H}}$ is increased from $1\times 10^{21}\mathrm{~cm}^{-2}$ to $4 \times 10^{21} \mathrm{~cm}^{-2}$, $a_{*}$ for SP2 varies from 0.931 to 0.834, with $\Delta a_{*}$ equaling 0.107. Because a higher value for $N_{\mathrm{H}}$ implies that the intrinsic disk emission must be larger, the luminosity increases while the disk temperature is much more weakly affected.  Accordingly, the inferred spin decreases with larger $N_{\mathrm{H}}$. In general, changing the value of $N_{\mathrm{H}}$ has a relatively small influence on the final spin outcomes when compared to the dynamical sources of uncertainty.


\begin{table*}
    \renewcommand\arraystretch{1.3}
    \centering
	\caption{When changing $N_{\mathrm{H}}$ for SP2, the best-fitting results with \texttt{crabcor*TBabs*(simpl*kerrbb2)} ~($\alpha=0.1$)}
    \begin{threeparttable}
    \begin{tabular}{ccccccc}
    \hline
    \hline
\texttt{TBabs} & \texttt{simpl}  &  \multicolumn{2}{c}{\texttt{kerrbb2}}& $\chi^2_\nu$ & $\chi^2$(d.o.f.) &$L/ L_{\mathrm{Edd}}$\\
\cline{3-4}
$N_{\mathrm{H}}$~($\times10^{22} \mathrm{cm}^{-2}$) &  $f_{\rm sc}$ & $a_{*}$ & $\dot{M} ~(\times 10^{18} \mathrm{~g} \mathrm{~s}^{-1})$ & &  \\
\hline
0.10   &  0.032  $\pm$ 0.010   &  0.931  $\pm$ 0.008   &  0.101  $\pm$ 0.005   &  0.942   &  200.550/213 &  0.014  \\ 
 0.18  &  0.037  $\pm$ 0.010   &  0.909  $\pm$ 0.009   &  0.117  $\pm$ 0.005   &  0.927   &  197.530/213 &  0.015\\   
0.30  &  0.045  $\pm$ 0.010   &  0.872  $\pm$ 0.011   &  0.144  $\pm$ 0.006   &  0.936   &  199.320/213 &  0.016 \\  
0.40   &  0.050  $\pm$ 0.009   &  0.834  $\pm$ 0.013   &  0.172  $\pm$ 0.007   &  0.969   &  206.330/213 &  0.018 \\                   
    \hline
    \end{tabular}
    \begin{tablenotes}
        \footnotesize   
  \item Notes. We display the following data: the hydrogen column density in units of $10^{22} \mathrm{cm}^{-2}$ ($N_{\mathrm{H}}$), the scattered fraction ($f_{\rm sc}$), the dimensionless spin parameter ($a_{*}$), the mass accretion rate of the disk in units of 10$^{18}$ g s $^{-1}$ ($\dot{M}$), the reduced chi-square ($\chi_{\nu}^{2}$), the total chi-square ($\chi^2$) and the degrees of freedom (d.o.f.), the bolometric Eddington-scaled luminosity ($L/ L_{\mathrm{Edd}}$). The errors are calculated in 68.3$\%$ confidence interval by \texttt{XSPEC}.  
    \end{tablenotes}
    \end{threeparttable}
    \label{table:6}
\end{table*}

\subsection{The influence of the photon index $\Gamma$}
\label{section:4.3}
To investigate the influence of $\Gamma$ on the spin, we tested $\Gamma$ by changing its value from 2.2 to 1.6, 1.8, 2.0, 2.4, 2.6, and 2.8. This is presented in Table \ref{table:7} for SP2 as representative of the ``gold" set. $a_{*}$ varies from 0.912 to 0.906 when $\Gamma$ is raised from 1.6 to 2.8, $\Delta a_{*}$ equaling 0.007. Ultimately, we find that the value of $\Gamma$ has a negligible influence on the final spin.

\subsection{The influence of viscosity parameter $\alpha $}
\label{section:4.4}
Although we present results for $\alpha $= 0.1 as our default in this work, we also complete the same computations for $\alpha = 0.01$. Table \ref{table:4} presents both sets of results, and we see that $\alpha = 0.01$ yields slightly larger values for the spin (for $\alpha = 0.01$, the average spin is 0.868 $\pm$ 0.039; for $\alpha = 0.1$, the average spin is 0.860 $\pm$ 0.039), as expected, consistent with findings in other systems, e.g., Cygnus X-1 (\citealt{Gou2011}; \citealt{Zhao2021}). We adopt an averaging over $\alpha=0.01$ and $\alpha=0.1$ for the final result.

\subsection{Discussion of the large variance in CF results}
\label{section:4.5}
In MAXI J1305-704, there is a very unusual phenomenon, that is, it has a high inner disk temperature ($\sim$0.9 keV) but a low Eddington fraction ($\sim$0.02 $L_{\mathrm{Edd}}$) (\citealt{Shidatsu2013}; \citealt{Morihana2013} similarly point out this anomaly). And at a given, fixed $M$, $i$, and $D$, the CF spin results show a substantially larger variance (0.74-0.95), which is anomalous vs other BHXRBs where that dispersion is very small compared to the spread introduced by the errors in $M$, $i$, and $D$. The typical variance in value is $\sim 3\%$ in radius, equivalent to $\Delta a_* \approx 0.05$ at $a_*=0$ or $\Delta a_* \approx 0.01$ at $a_*=0.9$. We speculatively suggest that these phenomena are coupled. The inner disk temperature of the \texttt{diskbb} component (see Table \ref{table:2}) is much higher than what is typical of a faint soft state in most BHXRBs with a low accretion rate. One possible reason is that the relativistic Doppler effects significantly modify the disk spectra due to the high inclination angle ($i={72_{-8}^{+5}}^{\circ}$; see \cite{Darias2013} to learn more). Or the high black hole spin significantly modifies the disk spectra. Meanwhile, due to the strong degeneracy between luminosity and spin, the large variance in CF results (which is an order of magnitude larger than usual) is plausibly related to its associated luminosity coupling and may also introduce noise into the continuum results for this system. We speculate that the strong and variable wind is at least partially responsible for the spin variations. The impact of disk winds on faint soft states for other high-inclination systems may be useful to assess this question in follow-up studies. 

The spin distribution of low-mass X-ray binaries (LMXBs) is relatively scattered, ranging from high spin to medium spin to low spin. The spin results of MAXI J1305-704 are in general agreement with the spin distribution observed in LMXBs (\citealt{Reynolds2021}; \citealt{Bambi2021}). In addition, \cite{steiner2012} reported a possible underlying link between black hole spin and mechanical jet power.


\begin{table*}
    \renewcommand\arraystretch{1.3}
    \centering
	\caption{When changing $\Gamma$ for SP2, the best-fitting results with \texttt{crabcor*TBabs*(simpl*kerrbb2)} ~($\alpha=0.1$)}
    \begin{threeparttable}
    \begin{tabular}{ccccccc}
    \hline
    \hline
 \multicolumn{2}{c}{\texttt{simpl}}  &  \multicolumn{2}{c}{\texttt{kerrbb2}}& $\chi^2_\nu$ & $\chi^2$(d.o.f.) &$L/ L_{\mathrm{Edd}}$\\
   \cline{1-2} \cline{3-4}
$\Gamma$ &  $f_{\rm sc}$ & $a_{*}$ & $\dot{M} ~(\times 10^{18} \mathrm{~g} \mathrm{~s}^{-1})$ & & & \\
\hline
1.6  &  0.024 $\pm$ 0.006     &  0.912 $\pm$ 0.008     &  0.116 $\pm$ 0.005     &  0.925   &  197.01/213&  0.015  \\   
1.8  &  0.027 $\pm$ 0.007     &  0.911 $\pm$ 0.008     &  0.116 $\pm$ 0.005     &  0.926   &  197.17/213&  0.015  \\   
2.0  &  0.032 $\pm$ 0.008     &  0.910 $\pm$ 0.009     &  0.116 $\pm$ 0.005     &  0.927   &  197.36/213&  0.015 \\    
2.2  &  0.037 $\pm$ 0.010     &  0.909 $\pm$ 0.009     &  0.117 $\pm$ 0.005     &  0.927   &  197.53/213&  0.015   \\  
2.4  &  0.045 $\pm$ 0.012     &  0.908 $\pm$ 0.009     &  0.117 $\pm$ 0.005     &  0.928   &  197.76/213&  0.015 \\    
2.6  &  0.053 $\pm$ 0.015     &  0.907 $\pm$ 0.010     &  0.117 $\pm$ 0.005     &  0.930   &  197.99/213&  0.015 \\    
2.8  &  0.063 $\pm$ 0.017     &  0.906 $\pm$ 0.010     &  0.118 $\pm$ 0.005     &  0.931   &  198.21/213&  0.015 \\
                 
    \hline
    \end{tabular}
    \begin{tablenotes}
        \footnotesize   
  \item Notes. Columns 4-9 show the dimensionless photon index ($\Gamma$), the scattered fraction ($f_{\rm sc}$), the dimensionless spin parameter ($a_{*}$), the mass accretion rate of the disk in units of 10$^{18}$ g s $^{-1}$ ($\dot{M}$), the reduced chi-square ($\chi^2_\nu$), the total chi-square ($\chi^2$) and the degrees of freedom (d.o.f.), the bolometric Eddington-scaled luminosity ($L/ L_{\mathrm{Edd}}$). The errors are calculated in 68.3$\%$ confidence interval by \texttt{XSPEC}.  
    \end{tablenotes}
    \end{threeparttable}
    \label{table:7}
\end{table*}

\section{CONCLUSIONS}
\label{section:5}
We perform a spectroscopic analysis for the black hole that resides in MAXI J1305-704, utilizing \emph{Swift}/XRT data in energy bands of 1.2-10.0 keV. We find that MAXI J1305-704 contains a moderate spin black hole with $a_{*}=0.87_{-0.13}^{+0.07}$ (1$\sigma$, averaging over $\alpha=0.1$ and $\alpha=0.01$) based on the model \texttt{kerrbb2} and the system parameters of \citealt{Mata2021}. Compared with other systems, MAXI J1305-704 is a relatively anomalous source with a larger-than-typical variance in CF spins, which is perhaps closely related to its unusually high-temperature disk at low ($\sim$2\% $L_{\mathrm{Edd}}$) luminosities. It should be noted that our spin constraint is limited by the precision of the dynamical measurements; more precise dynamical measurements would improve the CF spin constraint using these \emph{Swift}/XRT data.

\section*{Acknowledgements}
Y.F. thanks the sponsoring program in 2020 of the University of Chinese Academy of Sciences. Y.F. thanks Dr. Kim Page for the helpful suggestions in extracting \emph{Swift}/XRT spectra. Y.F. also appreciates Prof. M. Shidatsu, Prof. Timothy Kallman, and Prof. J.M.Miller for their helpful discussions on \texttt{XSTAR}. Y.F. thanks Dr. Yufeng Li for her suggestions on the manuscript. Also, Y.F. thanks Dr. Xi Long and Menglei Zhou for the insightful conversations. The \emph{Swift}/XRT data used in this study is provided by the UK Swift Science Data Centre at the University of Leicester in the United Kingdom. The software utilized is supplied by the High Energy Astrophysics Science Archive Research Centre (HEASARC), a service of NASA/ GSFC's Astrophysics Science Division and the Smithsonian Astrophysical Observatory's High Energy Astrophysics Division.

\section*{DATA AVAILABILITY}
This paper makes use of \emph{Swift}/XRT archival data which can be acquired from HEASARC: \\
\url{https://heasarc.gsfc.nasa.gov/cgi-bin/W3Browse/w3browse.pl}

\bibliographystyle{mnras}
\bibliography{ref}{}

\bsp	
\end{document}